\documentclass[aps,prb,twocolumn]{revtex4}
\usepackage{graphicx}% Include figure files
\usepackage{dcolumn}% Align table columns on decimal point
\usepackage{bm}% bold math
\usepackage{amsmath}
\def\one{\hbox{$1\hskip -1.2pt\vrule depth 0pt height 1.6ex width 0.7pt
                  \vrule depth 0pt height 0.3pt width 0.12em$}}
\begin{document}
\newcommand{\bk}{{\bf k}}
\newcommand{\bp}{{\bf p}}
\newcommand{\bv}{{\bf v}}
\newcommand{\bq}{{\bf q}}
\newcommand{\tbq}{\tilde{\bf q}}
\newcommand{\tq}{\tilde{q}}
\newcommand{\bQ}{{\bf Q}}
\newcommand{\br}{{\bf r}}
\newcommand{\bR}{{\bf R}}
\newcommand{\bB}{{\bf B}}
\newcommand{\bA}{{\bf A}}
\newcommand{\bK}{{\bf K}}
\newcommand{\vd}{{v_\Delta}}
\newcommand{\tr}{{\rm Tr}}
\newcommand{\kslash}{\not\!k}
\newcommand{\qslash}{\not\!q}
\newcommand{\pslash}{\not\!p}
\newcommand{\rslash}{\not\!r}
\newcommand{\bs}{{\bar\sigma}}

\title{Quasiparticle interference patterns as a test for the nature of the 
pseudogap phase in the cuprate superconductors}
\author{T. Pereg-Barnea$^{1,2}$ and M. Franz$^2$}
\affiliation{$^1$ Kavli Institute for Theoretical Physics, University of California, Santa Barbara, CA 93106-4030 \\
$^2$Department of Physics and Astronomy,
University of British Columbia, Vancouver, BC, Canada V6T 1Z1}
\date{\today}
\begin{abstract}
Electrons, when scattered by static random disorder, form standing waves 
that can be imaged using scanning tunneling microscopy. 
Such interference patterns, observable by the recently developed technique
of Fourier 
transform scanning tunneling spectroscopy (FT-STS), are shown to carry
unique fingerprints characteristic of the electronic order present in a 
material.
We exploit this feature of the FT-STS technique to propose a test for the 
nature of the enigmatic pseudogap phase in the 
high-$T_c$ cuprate superconductors.
Through their sensitivity to the quasiparticle spectra and coherence 
factors, the FT-STS patterns in principle carry enough information to 
unambiguously determine the nature of the condensate responsible for the 
pseudogap phenomenon. In practice, the absence of a detailed understanding 
of the scattering mechanism, together with the 
experimental uncertainties, prevent
such an unambiguous determination. We argue, however, that the next generation
of FT-STS experiments, currently underway, should be able to distinguish 
between the pseudogap dominated by the remnants of superconducting order from 
the pseudogap dominated by some competing order in the particle-hole channel.
Using general arguments and detailed numerical calculations, we point to
certain 
fundamental differences between the two scenarios and discuss the prospects
for future experiments.
\end{abstract}
\maketitle
%
%%%%%%%%%%%%%%%%%%%%%%%%%%%%%%%%%%%%%%%%%%%%%%%%%%%%%%%%%%%%%%%%%%%%%
\section{introduction}
The problem of high-$T_c$ superconductivity continues to engage the 
scientific community, yet much remains unknown.
The phase diagram of the cuprates is being elucidated with unprecedented 
accuracy but a unified
microscopic theory of the observed phases is still lacking.
While the question of how exactly the Mott-Hubbard antiferromagnetic 
insulator evolves into a high-$T_c$ superconductor upon hole doping remains 
unanswered, a number of phenomenological approaches to the problem have 
been developed. In the absence of a microscopic 
theory, progress can be made by means of a phenomenological description of 
the various phases, some of which are motivated
by microscopic arguments.\cite{emery1,balents1,so5,ft1,laughlin1,ddw1}

Among the various phases, two are especially well established: the 
antiferromagnet (AF) near the half filling and the superconducting phase which
occurs upon hole doping. The latter is reasonably well described by a 
BCS-like 
theory with a d-wave symmetric order parameter, $\Delta_\bk$.  It is this 
symmetry that gives rise to unique physics of the low energy excitations
which reside in the vicinity of the nodes of the gap function.

Intermediate between the AF and $d$-wave superconductor ($d$SC)
is the so called pseudogap phase which exhibits a gap in the single
particle density of states but is non-superconducting.\cite{timusk1} 
Experimentally, the pseudogap phase is characterized by a $d$-wave gap in the 
electron excitation spectrum,\cite{Sutherland} which  sets in at a 
temperature 
$T^*$ much higher then the superconducting transition temperature $T_c$.
The low energy nodal quasiparticles seem to exist in the pseudogap phase as 
well, however, they are strongly interacting and exhibit a high scattering 
rate.\cite{Valla}

Various authors have made proposals to describe the pseudogap phase
using different 
phenomenological approaches. One can identify two broad classes of 
theories.  One class includes theories of {\it competing orders}, in which 
the pseudogap results from the formation of static or fluctuating  order 
in the particle-hole channel. Examples include the 
AF spin density wave (SDW), the charge density wave (CDW), stripes, and 
the so-called $d$-density wave (DDW, also known as the flux phase). This 
order parameter is assumed to compete with  
superconductivity and its formation is energetically favorable 
in the pseudogap phase.\cite{so5,varma1,ddw1,vojta1}
The other class of theories follow the ideas of {\it order parameter phase 
fluctuations}, first introduced by Emery and Kivelson.\cite{emery1}  This 
school of thought  views the pseudogap as a superconductor on short length 
scales that fails to superconduct on longer length scales due to phase 
fluctuations.  In particular, this implies that 
a local superconducting order parameter is formed at $T^*$, but phase 
coherence is only achieved at $T_c$.\cite{emery1,randeria1,fm1,balents1,levin1,ft1,ftv1,laughlin1}
Unraveling the nature of this enigmatic phase is one of the central issues
in the physics of high-$T_c$ cuprates.

In this Paper we elaborate on a test for the nature of the pseudogap phase 
using the Fourier transform scanning tunneling spectroscopy (FT-STS), 
proposed by us earlier.\cite{pbf} We demonstrate
that the new generation of FT-STS experiments, performed in both the $d$SC 
and pseudogap phases,
 can provide a clear distinction between the two types of 
order, particle-hole (p-h)  or particle-particle (p-p), and can therefore 
serve as a referee between the above two classes of theories.  
Such a distinction is possible due to fundamental differences between the two
types of order. In a superconductor the elementary excitations consist
of an admixture of electrons and holes.  
The charge of these quasiparticles is not a good quantum number, and each 
quasiparticle carries a quantum phase that describes its particle-hole 
mixing.  As a result, scattering processes which give rise to the 
FT-STS signal, have high or low probability depending on the relative 
phase of the scattering quasiparticles.  This probability is given 
by the relevant {\it coherence factors}.
On the other hand, when the
scattering amplitude may depend on various microscopic details, we show that
there exist certain generic features which only depend on the type 
of ordering present in the system. 

The most straightforward test follows from the following simple
consideration. The interference patterns caused by a p-p order parameter are 
known; these are just the experimental results observed in cuprates below 
$T_c$.\cite{Hoffman,McElroy}
Comparing these patterns with the experimental results in the pseudogap 
phase should reveal the nature of the pseudogap order parameter.
If the patterns are similar to those of the $d$SC state, it is likely that 
the pseudogap order is in the p-p channel.  
If, on the other hand, an abrupt change in the patterns is found, the 
conclusion will be that a different type of order, in the p-h channel, 
governs the pseudogap. We shall discuss the relation to the available
experimental data in Sec. IV.C.

This article is organized as follows.  In the next section we 
revisit the formalism of the quasiparticle interference in the 
superconducting 
state, reviewing some recent theoretical work and commenting on the agreement 
with experiment.  We choose a framework for further analysis and discuss
how it may be justified on microscopic grounds. In section III we apply 
the same theoretical treatment to the pseudogap phase. We give a general
discussion of the problem and then study two representative theories of the 
pseudogap, the QED$_3$ theory and $d$-density wave (DDW).  We comment 
on their relevance in light of experimental observations.  Our conclusions
are given in section IV.

%%%%%%%%%%%%%%%%%%%%%%%%%%%%%%%%%%%%%%%%%%%%%%%%%%%%%%%%%%%%%%%%%%%%%%%%%%%%%%
\section{The superconducting state}  

\subsection{General considerations}
In the high resolution FT-STS experiments the spectrometer's sharp tip is 
placed above a freshly cleaved sample surface, and a bias voltage is applied 
between the sample and
tip. Tunneling occurs through the vacuum barrier and produces electric 
current which is measured as a function of the bias voltage.  
The measured differential conductance $dI(V,\br)/dV$ is proportional to the 
local density of states (LDOS)  $n(\br,\omega)$ of the sample below the STS 
tip at the point $\br$.  
A wide field of view of about 600\AA$\times$600\AA~is  scanned with atomic 
resolution, and at each bias voltage the spatial map of $n(\br,\omega)$ is 
collected. The spatial Fourier transform of this quantity, $n(\bk,\omega)$, is
then studied.
The resulting $k$-space distribution \cite{Hoffman,McElroy,Howald} reveals 
distinctive patterns with peaks at special wave vectors 
that correspond to certain periodic structures in the real space electron
wavefunctions. 
In a clean, homogeneous superconductor such structures would be absent; 
they apparently result from the 
translational symmetry breaking caused by disorder. At present, the 
details of this  scattering potential are not fully understood.

\begin{figure}
\includegraphics[width = 8.5cm]{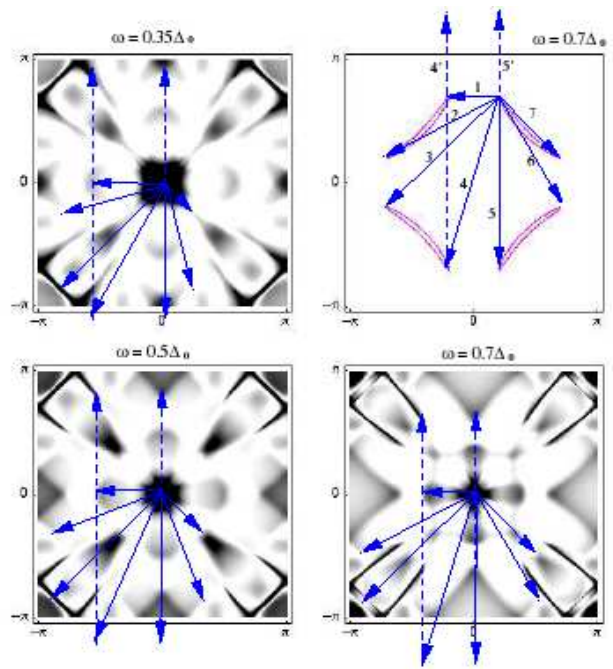}
\caption{Top right: The banana shape contours of constant low energy around 
the nodes of the gap and seven vectors connecting their ends at energy
 $\omega = 0.7\Delta_0$.\cite{Hoffman} 
Top left and bottom: Numerical evaluation of the Fourier transformed local 
density of states in Eq.~(\ref{eq:lam2}) with $t' = 0.3 t$, 
$\Delta_0 = 0.2t$, $\mu = -t$ and $\omega = 0.35$, $0.5$ and $0.7\Delta_0$ 
respectively. In order to best fit the experimental data\cite{Hoffman,McElroy} 
we present the real part of $\Lambda(q,\omega)$ and suppress the $\Delta_+\Delta_-$ 
term as explained in the text.
}\label{fig:dSC}
\end{figure}

These pioneering FT-STS experiments were followed by several theoretical 
studies that calculated the LDOS modulations in the BCS 
$d$-wave model using various approximations.\cite{Wang,Polkov,Capriotti,pbf}  
First, a heuristic picture referred to as the {\it octet model} was put 
forward by Hoffman {\it et al.} \cite{Hoffman,McElroy,Wang} and provided a 
simple 
and appealing explanation of the data.  The octet model identifies eight 
points in momentum space at which there is the largest single particle 
density of states (DOS), 
and asserts that the experimental FT LDOS is expected 
to have a peak at any point $\bq$ in the Brillouin zone such that $\bq$ 
connects any two of the octet points. As illustrated in Fig.~\ref{fig:dSC}
the latter occurs
at the ends of the banana shaped contours of constant energy (CCE), where 
the energy is equal to the bias voltage, $\hbar\omega=eV$.  

The original octet model explains the experimental data surprisingly well,
considering that it ignores several important issues.
In particular, the on-shell excitations away from the octet points are not 
taken into 
account.  One can argue that such excitations have lower single-particle DOS 
than the octet model points.  However, the ratio between the single particle 
DOS at the edges of the CCE bananas to the DOS of any other point on the CCE 
is only a result of the velocity anisotropy and is not as significant as other 
considerations.  This can be seen by expanding the energy dispersion near 
the nodes and calculating the DOS.  In the  ``nodal'' approximation,
scaling 
the momentum coordinates by the appropriate velocities ($v_F$ and $v_\Delta$) 
leads to a circular CCE, on which there is no special point that could 
define a peak position.  

A more detailed analysis compares the measured spectrum to a joint density of 
states (JDOS), $D(\bq,\omega)$, given by a $\bk$-space convolution of the 
single particle density of states
\begin{equation}\label{jdos}
D(\bq,\omega) = \sum_k \delta(\omega-E_\bk)\delta(\omega-E_{\bk-\bq}),
\end{equation}
where $E_\bk$ is the single particle excitation energy in the superconductor.
%a convolution of the single particle spectral functions 
%$A(\bk,\omega)=-\pi^{-1}{\rm Im}G_{11}(\bk,\omega)$,
%\begin{equation}
%D(\bq,\omega) = \sum_k A(\bk,\omega)A(\bk-\bq,\omega)
%\end{equation}
%where $G(\bk,\omega)$ is the electron propagator in the superconductor,
%to be specified shortly.
This approach includes the on-shell processes away from the nodal points
but neglects the off-shell processes as well as the coherence factors. 
We shall see shortly that in a $d$-wave superconductor the coherence
factors for scalar disorder are precisely such as to greatly enhance the
contributions from the octet points. We also find that the off-shell processes
mostly contribute to the background and, interestingly, tend to worsen the
agreement with experimental data.

It is interesting to note that the theoretically more satisfying approaches 
\cite{Wang,Polkov,Capriotti,pbf}, which model the LDOS using the full electron
Green's function (computed within various reasonable approximations), 
generally fail to account for the details of the experimentally observed 
patterns. The main problem is that these 
theoretical approaches typically produce
a number of prominent background features that are not present in experiment.
As a result, the computed patterns do not really resemble the experimental 
data (compare, e.g., Refs. \onlinecite{Wang} and \onlinecite{McElroy}), 
even if they contain
many of their characteristic features. At present it is not 
clear whether this discrepancy is a sign of some fundamental difficulty in 
the existing models or simply a question of details that have not been 
yet properly addressed. 

Here we take the latter view; i.e.\ that a simple BCS-type model of a $d$-wave 
superconductor with appropriate disorder should provide a complete description
of the FT-STS data once the details of the scattering mechanism have been
 ironed out.
In the present paper we are interested in these details only to the extent that
will allow us to establish a baseline for our discussion of FT-STS in the
{\em pseudogap} state. With this goal in mind in what follows we design a 
model for scattering in the $d$SC that produces patterns in reasonable
agreement with the experiment. We also explore several routes to justify this 
model from microscopic considerations, but in this we are only partially 
successful. We then take this same model and use it to discuss the pseudogap
state.

%-----------------------------------------------------------------------------
\subsection{The model}

We shall describe the system by a standard BCS $d$-wave Hamiltonian with 
disorder, treated as a perturbation. In the following, we consider three
types of disorder, namely in the charge, spin and pairing channels. 
We show how 
different types of disorder produce different scattering patterns and comment
on their relation to the patterns observed in the FT-STS experiments. Our
starting point is similar to that used by previous authors 
\cite{Wang,Polkov,Capriotti} but there will be some differences which will
allow us to closely reproduce the experimental data.

The system is described by the Hamiltonian 
${\cal H = H}_0 + \delta{\cal H}$, where
\begin{equation}
\nonumber
{\cal H}_0 = \sum_\bk \psi^\dagger_\bk
\begin{pmatrix}
        \epsilon_\bk\,~~~\Delta_\bk\\ \Delta^*_\bk\, ~-\epsilon_\bk
\end{pmatrix} 
\psi_\bk 
\end{equation}
is the $d$-wave BCS Hamiltonian describing the pure, uniform $d$SC.
$\psi^\dagger_\bk = (c_{\bk\uparrow}^\dagger,
c_{-\bk\downarrow})$ is the Nambu space creation 
operator and $\epsilon_\bk$, $\Delta_\bk$ are the band structure and gap 
function respectively. In what follows, we choose to work with the 
second-nearest neighbor hopping model and a $d_{x^2-y^2}$ gap function
\begin{eqnarray}
\nonumber
\epsilon_\bk &=& -2t(\cos k_x+\cos k_y)-4t'\cos k_x\cos k_y -\mu,\\
\Delta_\bk &=& {1\over 2}\Delta_0(\cos k_x -\cos k_y).
\label{disp}
\end{eqnarray}
Disorder is generically described by a Hamiltonian
\begin{equation}
\delta{\cal H}=\sum_{i,j,\sigma,\sigma'} V_{ij}^{\sigma\sigma'} 
c^\dagger_{i\sigma} c_{j\sigma'}^{}
=\sum_{\bk,\bk'}\psi^\dagger_\bk \hat{V}_{\bk\bk'} \psi_{\bk'}.
\end{equation}
Here $i,j$ is a lattice index, $V_{ij}^{\sigma\sigma'}$ is the random potential
and  $\hat{V}_{\bk\bk'}$ its Fourier transform, a $2\times 2$ matrix in
Nambu space. 

The electron propagator of the perturbed system can be expressed as 
\begin{equation}\label{t}
G(\bk,\bk',\omega) = G^0(\bk,\omega)\delta_{\bk,\bk'}+G^0(\bk,\omega)
\hat{T}_{\bk\bk'}(\omega)G^0(\bk',\omega),
\end{equation}
where
\begin{equation}\label{eq:dBCS}
G^0(\bk,\omega) = [\omega -\sigma_3\epsilon_\bk-\sigma_1\Delta_\bk]^{-1}
\end{equation} 
is the bare Green's function and $\sigma_i$ are the Pauli matrices.
$\hat{T}_{\bk\bk'}(\omega)$ is the $T$-matrix,
subject to the Lippman-Schwinger equation
\begin{equation}\label{t1}
\hat{T}_{\bk\bk'}(\omega) = \hat{V}_{\bk\bk'} + \sum_\bq 
\hat{V}_{\bk\bq} G^0(\bq,\omega)\hat{T}_{\bq\bk'}(\omega). 
\end{equation}
The local density of states at the point $\br$ of the sample is then given by
\begin{equation}\label{ldos1}
n(\br,\omega)=-{1\over \pi}{\rm Im}[G_{11}(\br,\br,\omega)+
G_{22}(\br,\br,-\omega)],
\end{equation}
where $G(\br,\br',\omega)$ is the spatial Fourier transform of 
$G(\bk,\bk',\omega)$. Here and hereafter ``Im'' is understood as taking 
the usual difference between the retarded and the advanced quantity 
measuring the discontinuity of the propagator across the real
frequency axis. It  is important to keep this in mind since 
$\hat{V}_{\bk\bk'}$ is in general complex but its 
imaginary part does {\em not} contribute to the above discontinuity and
therefore to LDOS in Eq.~(\ref{ldos1}). 
Taking the ``Im'' symbol literally would lead to incorrect results.

For arbitrary disorder one must typically solve for the $T$-matrix from 
Eq.~(\ref{t1}) using numerical methods. We shall not attempt such a 
detailed numerical solution here. 
Rather, we shall limit ourselves to two different approximations where 
progress can be made analytically. First, we consider the limit of {\em weak}
but otherwise arbitrary disorder. Second, we study the case of arbitrarily 
strong but {\em dilute} point-like scatterers.

%----------------------------------------------------------------
\subsection{Weak scattering (Born limit)}

To leading order in the scattering potential $V$, i.e., in the Born 
approximation, we have $\hat{T}_{\bk\bk'}(\omega)\simeq \hat{V}_{\bk\bk'}$ and
the perturbed Green's function is given by
\begin{equation}\label{eq:born}
G(\bk,\bk',\omega) = G^0(\bk,\omega)\delta_{\bk,\bk'}+G^0(\bk,\omega)
\hat{V}_{\bk\bk'}G^0(\bk',\omega).
\end{equation}
Since the first term of Eq.~(\ref{eq:born}) is uniform, all the interesting 
information resides in the second term. 

We now focus on the disorder potential in the {\em charge} channel. In this 
case we have on-site random potential $V_i$ coupled to electron charge,
\begin{equation}
\delta{\cal H}=\sum_{i,\sigma} V_{i} 
c^\dagger_{i\sigma} c_{i\sigma}^{}
=\sum_{\bk,\bk'}V_{\bk-\bk'}\psi^\dagger_\bk\sigma_3 \psi_{\bk'}
\end{equation}
i.e., $\hat{V}_{\bk\bk'}=\sigma_3V_{\bk-\bk'}$, where $V_\bq$ is the Fourier
transform of $V_i$.
The advantage of the Born limit is that one may express 
the interesting non-uniform part of the measured FT LDOS, 
$\delta n(\bq,\omega)$, in a factorized form \cite{Capriotti}
\begin{eqnarray}\label{eq:components}
\delta n(\bq,\omega) =-{1\over \pi} |V_\bq|  
{\rm Im} \left[\Lambda_{11}(\bq,\omega)+ \Lambda_{22}(\bq,-\omega) \right],
\end{eqnarray}
where
\begin{equation}\label{eq:lambda}
\Lambda(\bq,\omega) = \sum_\bk G^0(\bk,\omega)\sigma_3
G^0(\bk-\bq,\omega).
\end{equation}
As discussed by Capriotti {\em et al.} $V_\bq$ is a random function with no
interesting structure while $\Lambda(\bq,\omega)$ contains information about
the underlying clean system. Any peaks observed in $\delta n(\bq,\omega)$
therefore must be attributed to $\Lambda(\bq,\omega)$. 
Thus, by studying the interference patterns in a disordered
sample one can learn about the properties of the system in the absence of 
disorder.

\begin{figure}
\includegraphics[width=7.8cm]{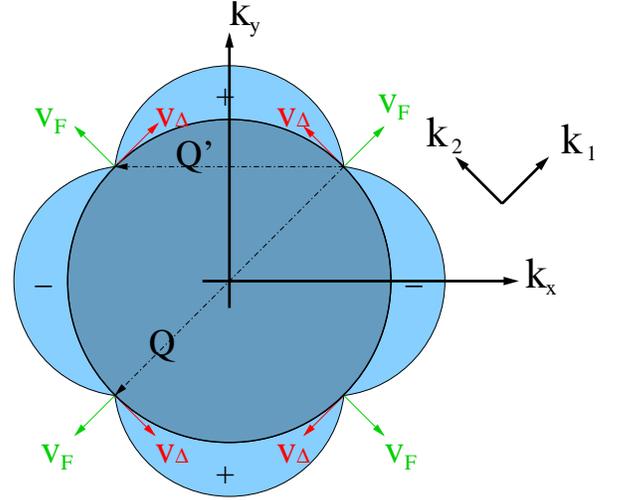}
\caption{The nodes of the $d$SC gap on the underlying Fermi surface with the velocities $v_F$ and $v_\Delta$
as defined in each node. The vectors $\bQ$ and $\bQ'$ span antipodal and adjacent nodes respectively.}
\label{fig:nodes}
\end{figure}

\begin{figure}
\includegraphics[width = 8.5cm]{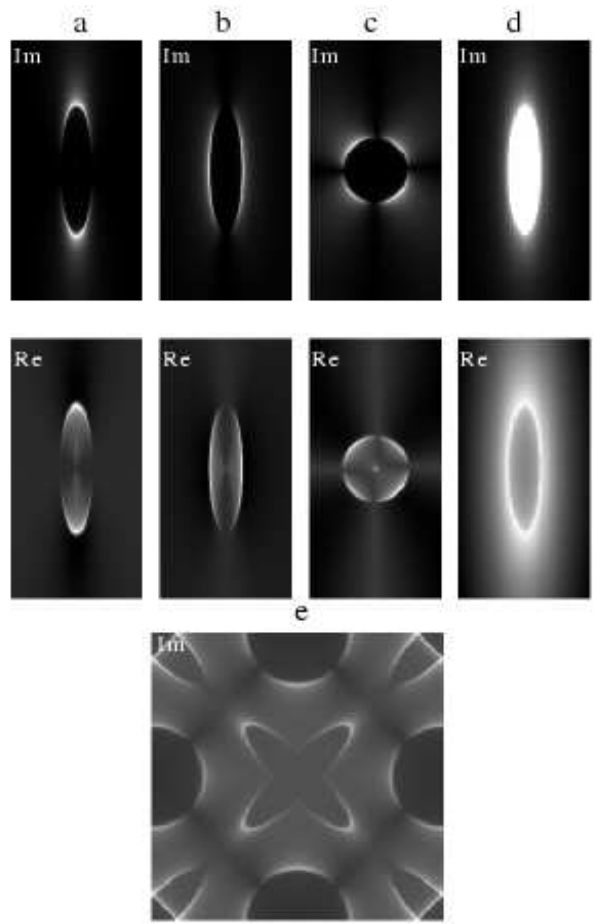}
\caption{Interference patterns in the nodal approximation. a)
Non-magnetic scattering within a single node Eq.~(\ref{eq:born}), b) 
Non-magnetic scattering between two antipodal nodes, Eq.~(\ref{eq:lin34}), 
c) Magnetic scattering between adjacent nodes,  Eq.~(\ref{eq:lin12}), 
d) Magnetic scattering within a single node, or antipodal nodes and non 
magnetic scattering of adjacent nodes.
In (a,b,d) we used $v_F/v_\Delta=4$ and in (c) $v_F/v_\Delta=1$. 
Note that the axes here represent the nodal coordinates,
$q_1$ (horizontal) and $q_2$.
 e) A composite of the intra- and antipodal-node terms for non-magnetic scattering and adjacent-node magnetic scattering showing the full Brillouin zone distribution in the nodal approximation.
}
\label{fig:lin}
\end{figure}

With the $d$-wave BCS Green's function, Eq.~(\ref{eq:dBCS}), one finds
\begin{equation}\label{eq:lam2}
\Lambda_{11}(\bq,i\omega) = {1\over L^2}\sum_\bk 
{(i\omega+\epsilon_+)(i\omega+\epsilon_-)-\Delta_+\Delta_-\over
(\omega^2+E_+^2)(\omega^2+E_-^2)},
\end{equation}
with $\epsilon_\pm=\epsilon_{\bk\pm\bq/2}$, 
$\Delta_\pm=\Delta_{\bk\pm\bq/2}$ and 
$E_\pm=\sqrt{\epsilon_\pm^2+\Delta_\pm^2}$.
Eq.~(\ref{eq:lam2}) can be evaluated exactly using numerical methods.  
However, before 
doing so we shall elucidate its main features by employing the 
nodal approximation.  In this approximation, valid at energies low compared
to the 
maximum gap $\Delta_0$, the spectrum is linearized in the vicinity of the 
nodes, and we consider scattering processes either within the same node or 
between different nodes.
Let us first evaluate the LDOS resulting from intra-node scattering, 
which corresponds to the octet vector $\bq_7$.\cite{McElroy}
For this case, Eq.~(\ref{eq:lam2}) can be approximated in the following way:  
near the node we use the $45^{\rm o}$ rotated coordinates $k_1$ and $k_2$ 
shown in Fig.~(\ref{fig:nodes}).
To first order in the momentum we expand $\epsilon_\bk \to v_F k_1$ and 
$\Delta_\bk \to v_\Delta k_2$.
We then rescale the momentum coordinates so that 
$v_F k_1\to k_1$, $v_\Delta k_2 \to k_2$, and redefine 
${1\over 2} v_F q_1\to \tq_1$ and ${1\over 2} v_\Delta q_2\to \tq_2$.  
In the scaled coordinate frame $E_\bk = |\bk|$ and Eq.~(\ref{eq:lam2}) 
simplifies to
\begin{equation}\label{eq:lin7}
\Lambda_{\rm lin} = {1\over v_Fv_\Delta}\int {d^2k\over(2\pi)^2} 
{-\omega^2 + (k_1^2-k_2^2)-(\tq_1^2-\tq_2^2)\over
[\omega^2+(\bk+\tbq)^2][\omega^2+(\bk-\tbq)^2]}.
\end{equation}
As shown in appendix \ref{inter-nodal}, the explicit solution can be 
found using the Feynman parameterization\cite{peskin} and reads:
\begin{eqnarray}\label{eq:lin7s}
\Lambda_{\rm lin}(\bq,\omega) &=&{1\over 2\pi v_Fv_\Delta}
\left[\left({\tq_2\over\tq}\right)^2 
{\cal F}\left({\omega\over\tq}\right)
-{1\over 2}\right],\nonumber\\
{\cal F}(z)&=&1-{z^2\over\sqrt{z^2-1}}\arctan{1\over\sqrt{z^2-1}}.
\end{eqnarray}
The function  ${\cal F}(z)$ implies an inverse square root singularity in both the real and
imaginary parts of $\Lambda_{\rm lin}(\bq,\omega)$ along an elliptic 
contour of constant energy given by $E_\bq=2\omega$ (here
$E_\bq=2\tq\equiv\sqrt{v_F^2q_1^2+v_\Delta^2q_2^2}$). More importantly,
this singularity is weighted by an angular factor $(\tq_2/\tq)^2=
(v_\Delta q_2/E_\bq)^2$, producing the largest amplitude at the two ends
of the ellipse, as illustrated in Fig.~(\ref{fig:lin}a). These points
of largest intensity coincide with $\pm\bq_7$. It is important to note
that the angular dependence results purely from the coherence factors
(i.e.\ terms in the numerator of  Eq.~\ref{eq:lin7}) and would not appear
in the JDOS Eq.~(\ref{jdos}). 

This particular angular dependence can be understood by analyzing
the relevant BCS coherence factors as follows. Consider first 
$\bq$ that connects the two CCE point along the nodal direction. The 
process represents scattering of a pure particle to a pure hole 
since along this direction there is no p-h mixing due to the vanishing gap.  
The amplitude of such a process is zero due to the charge coupling.
On the other hand the two points at the edges of the CCE both have equal 
particle and hole mixing
($|u_\bk|=|v_\bk|$ in the BCS notation) and the same quantum phase.  
This leads to constructive interference and maximal $\Lambda(\bq,\omega)$.

Similarly, we can treat magnetic disorder. In this case, the potential 
couples to the electron spin and is therefore proportional to the unit matrix 
$\sigma_0$=\one \ in Nambu space:
\begin{equation}
\delta{\cal H} = \sum_i V_i^{\rm mag}[c^\dagger_{i\uparrow} c^{}_{i\uparrow} 
- c^{\dagger}_{i\downarrow} 
c^{}_{i\downarrow}] = 
\sum_{\bk,\bk'}V_{\bk-\bk'}^{\rm mag}\psi^\dagger_\bk\sigma_0\psi_{\bk'}^{}.
\end{equation}
Again, one can describe the scattering patterns using $\Lambda(\bq,\omega)$. 
It is easy to see that the expression for this quantity is just like 
Eq.~(\ref{eq:lam2}) except for the sign change in front of the 
$\Delta_+\Delta_-$ term.  
In the nodal approximation for intra-nodal scattering we thus get
\begin{equation}\label{eq:lam5}
\Lambda_{\rm lin}^{\rm mag} = {1\over v_Fv_\Delta}\int {d^2 k\over(2\pi)^2} 
{-\omega^2 + \bk^2-\tbq^2 \over
[\omega^2+(\bk+\tbq)^2][\omega^2+(\bk-\tbq)^2]}.
\end{equation}
The solution (see Appendix A) is now very different,
\begin{eqnarray}\label{eq:lam5s}
\Lambda_{\rm lin}^{\rm mag}(\bq,\omega) &=&{-1\over 2\pi v_Fv_\Delta}
\left[i\pi + 2 {\cal F}'\left({\omega\over\tq}\right)
+\ln\left({\omega^2\over\lambda^2}\right)\right],\nonumber \\
{\cal F}'(z)&=&\sqrt{z^2-1}\arctan{1\over\sqrt{z^2-1}}.
\end{eqnarray}
Here, since the spin is a good quantum number, no angular dependent coherence 
factors are present, and the interference amplitude is constant along 
the elliptic contour $2\omega=\sqrt{v_F^2q_1^2+v_\Delta^2q_2^2}$, as
shown in Fig.~(\ref{fig:lin}d). The square root singularity 
is replaced by a cusp along the contour. This result provides another
example of the importance of the coherence factors.

The linearized treatment for internodal magnetic and non-magnetic scattering 
is similar and given in appendix~\ref{inter-nodal}. A composite picture 
including scattering processes
between all nodes is shown in panel (e) of Fig.~(\ref{fig:lin}). We observe 
that it indeed contains peaks and qualitatively exhibits features observed
in FT-STS experiments.
The linearized treatment also indicates that non-magnetic scattering alone 
cannot produce all of the experimentally observed features.  The octet points 
$\bq_1$,  $\bq_5$ and $\bq_{2,6}$, corresponding to adjacent nodes 
scattering, can all be produced together only by including magnetic scattering 
(see Fig.~(\ref{fig:lin}c)).  This combination of magnetic and non-magnetic 
scattering is discussed in the next subsection.
\begin{figure}
\includegraphics[width= 8.5cm]{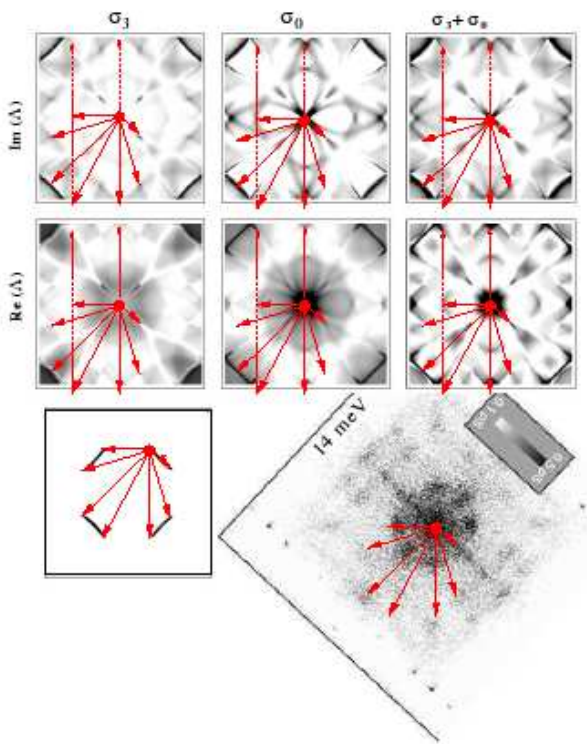}
\caption{Top six panels: numerical evaluation of $\Lambda(\bq,\omega)$ for 
$\omega = 0.32\Delta_0$, where the scattering potential is, from left to 
right  proportional to $\sigma_3$ (non-magnetic scattering), $\sigma_0=\one$ 
(magnetic scattering) and an equal combination of both.
Bottom: The octet points at the edges of the banana shaped CCE (left) and 
the experimental data of 14meV$\approx 0.32\Delta_0$ from 
Ref.~\onlinecite{McElroy} 
(right).  Note that the Brillouin zone of the experimental data is 45$^{\rm o}$ 
rotated with respect to the crystallographic directions.}
\label{fig:schemes}
\end{figure}

Beyond the linearized approximation, the 
interference patterns can be evaluated exactly, by numerical methods, using 
the full model dispersion Eq.~(\ref{disp}). Work along these lines has
been done for a single point like scatterer in Refs.\ \onlinecite{Wang,Polkov,
Capriotti} and for arbitrary weak disorder in 
Refs.\ \onlinecite{Capriotti,Zhu}. 
To evaluate Eq.~(\ref{eq:lam2}) numerically we find it 
convenient to perform the momentum space Green's function convolutions by 
employing the fast Fourier transform (FFT) technique. 
In particular we transform $G^0(\bk,\omega)$ to real space, multiply the 
appropriate components, and perform an inverse FFT back to the momentum 
space. Since the run time of FFT is only $o(n\log(n))$ where $n$ is the 
array size, this allows for fast evaluation of $\Lambda(\bq,\omega)$ on large 
lattices (of up to $1024\times 1024$ lattice points).

The comparison of numerical results with the experimental data shown in 
Fig.~(\ref{fig:schemes}) leads to two interesting insights.  
When considering the Born limit of point-like charge/spin impurities 
represented by $\sigma_3$ or $\sigma_0$ matrices respectively, it appears 
that the  real rather then the imaginary part of $\Lambda(\bq,\omega)$ 
is describing the experiment more accurately.  In addition, one achieves even 
better agreement with the data when neglecting the off-diagonal part of the 
unperturbed Green's function [i.e.\ $\Delta_+\Delta_-$ term in 
Eq.~(\ref{eq:lam2})].  
Operationally, such suppression of the off diagonal Green's 
function may be achieved as a result of an even combination of magnetic and 
non-magnetic scattering since the two contain contributions proportional
to $\Delta_+\Delta_-$ with opposite signs. Fig.~(\ref{fig:schemes}) shows our 
numerical 
calculation of $\Lambda(\bq,\omega)$ for a magnetic, non-magnetic and an 
even combination of both types of scattering. Our best fit to the data is 
obtained by taking the {\em real} part of the equal mixture of magnetic and 
non-magnetic scattering  and is  presented  in Fig.~(\ref{fig:dSC}).

We adopt this result for FT-STS in $d$SC as our baseline for studies of 
the pseudogap state. Before we proceed to study the latter we briefly discuss
how one could justify this model based on the physics beyond the Born 
approximation.

%-----------------------------------------------------------------------------
\subsection{Beyond the Born limit: the $T$-matrix}

As mentioned above, the Born limit scattering model is in best agreement with 
the experimental data when (i) the off-diagonal part of the bare Green's 
function, ${\cal F}(\bk,i\omega)=\Delta_\bk /(\omega^2+E_\bk^2)$,
is  ignored and (ii) the real part of $\Lambda$ rather then the imaginary 
part is 
taken (in contrast to the prescription in Eq.~(\ref{eq:components})). 
The question naturally arises: can this prescription be justified in a 
physically well motivated manner?

(i) may arise simply as a result of chemistry; it is in principle possible
that impurities in cuprates act both as charge and spin scatterers and that
the strength in both channels is roughly equal. However, such an assumption
amounts to a fine tuning of parameters. Furthermore, chemistry
cannot plausibly explain (ii).
 
We now show that at least qualitatively both effects may arise when the 
disorder 
potential is not weak and the full $T$-matrix is considered. 
In the $T$-matrix, all of the terms in the Born series are resummed by means 
of  Eq.~(\ref{t1}). Unfortunately, solving the Lippman-Schwinger equation
for arbitrary disorder potential is a daunting task. Progress can be made,
however, in the case of point-like scaterrers. In particular, for a single
point-like scatterer at the origin, the $T$-matrix becomes momentum 
independent and its $\omega$ dependence can be found 
analytically.\cite{Balatsky1} One can easily 
see that the expression for $\delta n(\bq,\omega)$ must be modified as
\begin{equation}\label{lam3}
\delta n(\bq,\omega) =-{1\over \pi}{\rm Im} 
\sum_\bk G^0(\bk,\omega)\hat{T}(\omega)G^0(\bk-\bq,\omega).
\end{equation}
In the following,
we focus on this analytically tractable situation which is also relevant to
a dilute concentration of point scatterers provided that one can neglect 
coherent scattering from multiple impurities. 

First note that in a superconductor, a charge scatterer, which is represented 
by a potential proportional to $\sigma_3$ in Nambu space, generates terms
proportional to $\sigma_0$ in the Born series. These terms are {\it 
magnetic-like} since the magnetic potential is represented by a unit matrix.  
Magnetic-like terms arise from a pure charge impurity due to the $2\times 2$ 
structure of the Green's function (there is no symmetry that forbids their 
creation).  Second, the $T$-matrix is in general complex. Thus, there could
be a regime in which its imaginary part dominates, implying that 
$\delta n(\bq,\omega)$ would receive dominant contributions from the real part
of $\Lambda(\bq,\omega)$.

As shown by Balatsky {\it et al.},\cite{Balatsky1} near 
half filling summing the Born series for the point-like charge impurity 
yields
$\hat{T}(\omega) = T_0(\omega)\sigma_0 + T_3(\omega)\sigma_3$
with
\begin{equation}
T_0(\omega) = {g_0(\omega)  \over c^2 - g_0^2(\omega)}, \;\;\;\;\;\;\; 
T_3(\omega) = {-c \over c^2 - g_0^2(\omega)}
\end{equation}
where $c=\cot \delta_0$ is the cotangent of the 
scattering phase shift and $g_0(\omega) = \tr\sum_\bk G^0(\bk,\omega)$.
The above frequency structure of the $T$-matrix represents a quasi-bound 
state near the impurity with a
resonance at $c = \pm g_0(\omega)$.  Since $c$ is a real number and the 
Green's function has an 
imaginary part, at resonance the $T$-matrix is predominantly imaginary and 
$T_3(\omega)\approx \mp T_0(\omega)$.  This means that near resonance both
(i) and (ii) can be satisfied and the prescription that we obtained 
heuristically in the previous subsection can be physically motivated.

Our justification above should be viewed merely 
as a ``proof of principle'' since
the arguments only hold near the resonant frequency whereas we need them
to apply at all frequencies. One could perhaps argue that in real samples 
the impurity resonances will be distributed forming a continuous band in 
which $\hat{T}(\omega)$ is predominantly imaginary and $T_3(\omega)\approx 
T_0(\omega)$. However at this stage this is merely a speculation.

Another mechanism that can give rise to suppression of the off-diagonal 
terms is spatial variation of the gap amplitude. Such variations have indeed
been observed  in the STS studies on BiSCCO samples.\cite{Pan1,Wang2} 
Theoretical treatment of these is somewhat complicated by the fact that 
in a $d$SC the order parameter lives on the nearest neighbor bonds of the
underlying ionic lattice. For that
reason we defer this discussion to the Appendix \ref{bond}.

\begin{figure}
\includegraphics[width = 8.5cm]{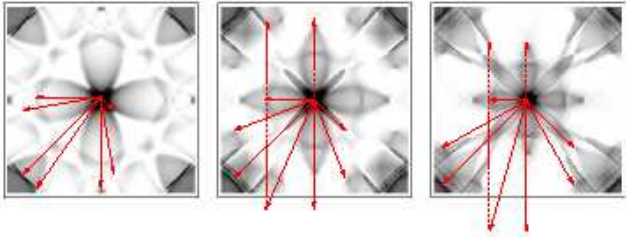}
\caption{The $T$-matrix approximation for a point like charge impurity 
and a local gap modulation at the origin.  The band structure parameters and 
gap magnitude are the same as 
in Fig.~(\ref{fig:dSC}) and $V=0.3t,\delta\Delta = 0.3\Delta_0$. 
The frequencies from left to right 
are $0.2,0.5$ and $0.7\Delta_0$.}\label{fig:t-matrix}
\end{figure}

Our numerical results for a point-like impurity (including the gap 
suppression) in the $T$-matrix approximation
at the relevant probing frequencies are presented in 
Fig.~\ref{fig:t-matrix}.
The obtained results are similar to those obtained by the simple treatment 
of section IIC (Fig.~\ref{fig:dSC}) and display most of the features 
expected by the octet model.
We emphasize again that considerations in this subsection are not meant as
detailed modeling of the scattering potentials in the cuprates. Rather, they
are meant to illustrate how the heuristic treatment presented in the previous 
subsection can be reconciled with a more microscopic approach. In what follows
we shall continue to describe the FT-STS patterns $\delta n(\bq,\omega)$  in 
terms of $\Lambda(\bq,\omega)$ via Eq.~(\ref{eq:components}).

\section{The Pseudogap State}

We now focus on the pseudogap state. In principle, the information contained
in the FT-STS patterns should permit the unambiguous determination of the 
electron order present in underdoped cuprates above $T_c$ (but below $T^*$).
In practice, however, we have seen that details of the scattering mechanism
can lead to vastly different interference patterns, and even the 
description of
the well-understood $d$SC state is fraught with difficulty. We thus set our
sights on a more modest goal; in the following we consider the question
whether it is possible, using the FT-STS techniques, to identify the 
condensate in the pseudogap state as being in the p-p (superconducting) or p-h
channel. 

As we have seen above the p-p order in the superconducting phase gives
rise to very special BCS coherence factors which mix electron and hole degrees
of freedom. These coherence factors lead to unique FT-STS patterns with sharp
peaks at the set of ``octet vectors'' shown in Fig.~(\ref{fig:dSC}). According
to our analysis there are two crucial factors which determine the position of 
these peaks in the Brillouin zone. The quasiparticle {\em excitation spectrum}
$E_\bq$ defines a set of contours of constant energy, $E_\bq=\pm\omega$,
along which the denominators of Eq.~(\ref{eq:lam2}) vanish and cause large
FT-STS response. The {\em coherence factors}, entering
the numerator of Eq.~(\ref{eq:lam2}), determine whether the quasiparticle
interference is constructive or destructive at a given point 
and thus select special points along these contours at which the peaks in 
FT-STS appear.

It follows from this analysis that the observed FT-STS patterns will 
change when
either the excitation spectrum {\em or} the coherence factors are modified as 
one passes from $d$SC to the pseudogap state. This observation is the key to 
our proposal to distinguish between the pairing and p-h correlations in 
the pseudogap state. If, on crossing the $T_c$, the FT-STS patterns are 
essentially unmodified, then one can conclude that the pseudogap is dominated
by ordering in the p-p channel. If, on the other hand, a qualitatively 
different pattern is observed above $T_c$, 
then this would be evidence for another type
of order, most likely in the p-h channel. These conclusions should be 
insensitive to the details of the scattering mechanism that give rise to 
the quasiparticle interference.  

A more complicated situation may arise when the competing
order can coexist with superconductivity in some region of the phase 
diagram, as envisioned e.g.\ in Ref.\ \onlinecite{ddw1}. In that case 
the above analysis still applies but one expects a more gradual change of
the interference patterns as a function of temperature (or another control 
parameter such as doping) reflecting the gradual growth of the competing
order at the expense of superconductivity.

In order to illustrate the crucial role the type of order parameter plays in 
the process of quasiparticle scattering we examine two representative models 
of the pseudogap. Our goal is to understand how the interference patterns
vary depending on whether the order parameter is in the p-p channel 
(superconductivity) or in the particle-hole channel.

%--------------------------------------------------------------------------
\subsection{Superconducting phase fluctuations}

In this scenario the pseudogap state is viewed as a phase-disordered $d$-wave
superconductor.\cite{emery1} There exist several theoretical frameworks
to describe such a state of electronic matter
\cite{balents1,randeria1,fm1,levin1,ft1} all sharing a common feature that 
the underlying order is in the pairing
channel. Here we focus on the ``QED$_3$'' theory\cite{ft1,ftv1} but it
is easy to see that other approaches will yield similar general outcomes
for the FT-STS patterns.

In the QED$_3$ theory, the phase of the superconducting order parameter
is disordered by vortex-antivortex excitations that are encoded in a U(1) 
gauge field minimally coupled to the nodal fermions.  
At $T_c$, vortex-antivortex pairs unbind through the Kosterlitz-Thouless 
transition and cause
a  loss of phase coherence at long distances.  The resulting 
pseudogap state is described by nodal fermions that are strongly 
interacting, the interaction mediated by the massless U(1)
gauge field.  The fermionic Green's function in Nambu space\cite{ftv1} is 
given by
\begin{equation}\label{eq:luttinger}
G^0(\bk,i\omega)= \lambda^{-\eta}{i\omega + \epsilon_\bk\sigma_3 
\over [\omega^2+\epsilon_\bk^2+\Delta_\bk^2]^{1-\eta/ 2}},
\end{equation}
where $\lambda$ is a high energy cutoff and $\eta$ is the anomalous dimension
exponent which encodes the dressing of the nodal quasiparticles by the 
gauge field fluctuations. General requirements of causality and unitarity 
dictate that $\eta\geq 0$; however the exact value of the anomalous 
dimension is still under debate.\cite{ftv1,khvesh2,rantner2,gusynin1} Most 
believe that $\eta$ is a small positive number. Here we sidestep these  
issues by treating $\eta$ as a free parameter. We show that the FT-STS patterns
are insensitive to its exact value.

\begin{figure}
\includegraphics[width = 8.5cm]{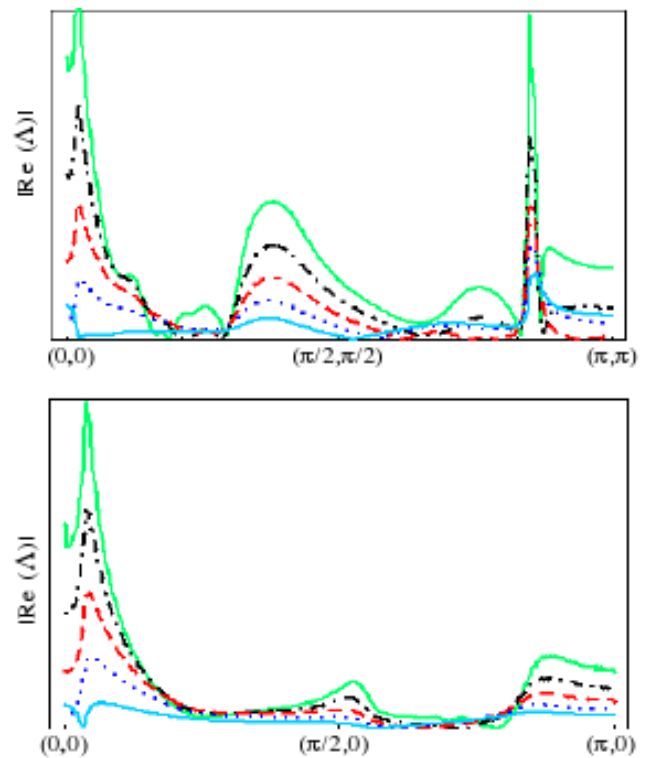}
\caption {The interference patterns within the QED$_3$ scenario 
with the anomalous dimension exponent  $\eta = 0, 0.1,0.2,0.3$ and $0.4$, 
$\omega = 0.45\Delta_0$ and all other parameters as in Fig.~(\ref{fig:dSC}). 
The upper panel displays a cut along the $(\pi,\pi)$ direction of the 
Brillouin zone and the bottom plot is a cut along the $(\pi,0)$ direction.
In both panels, the solid, dashed-dotted, dashed, dotted and solid lines
correspond to $\eta = 0, ..., 0.4$ respectively.}
\label{fig:eta}
\end{figure}

The similarity between QED$_3$ propagator (\ref{eq:luttinger}) and that of
the $d$SC state suggests that the interference patterns obtained in this model 
will resemble those in the $d$SC state. 
We study the FT-STS patterns within the same set of approximations adopted
for the $d$SC state, i.e. we evaluate $\Lambda(\bq,\omega)$ and consider 
its real part. In addition, we note that 
the absence of the off-diagonal part in the  QED$_3$ propagator (which reflects
lack of long range superconducting order in the pseudogap state) implies 
that the $\Delta_+\Delta_-$ term in $\Lambda(\bq,\omega)$ will automatically 
vanish, irrespective of the type of scattering. 
 Numerical evaluations of $\Lambda(\bq,\omega)$ are carried out using 
the fast Fourier transform technique described above and the 
results are given in Fig.~(\ref{fig:eta}). As expected based on our general
argument the QED$_3$ interference patterns are similar 
to those obtained for the $d$SC state.  Peaks occur at the same points in the 
Brillouin zone and the anomalous dimension $\eta$ causes some suppression of 
the sharp singularities.

This similarity can also be understood by performing the nodal approximation 
for the QED$_3$ propagators, which is done in appendix~\ref{inter-nodal}.  
This treatment shows that the coherence factors produce the same angular 
dependence along the elliptic contour as in the $d$SC state, but the
square root singularity is replaced by a weaker $1/(z^2-1)^{1/2-\eta}$ 
singularity, where $z= \omega/|\tbq|$ as before. Thus, the QED$_3$ and $d$SC
patterns will be qualitatively similar as long as $0\leq\eta< {1\over 2}$, 
which is exactly the range of values where we expect $\eta$ to lie based
on general considerations.
\begin{figure}
\includegraphics[width = 7cm]{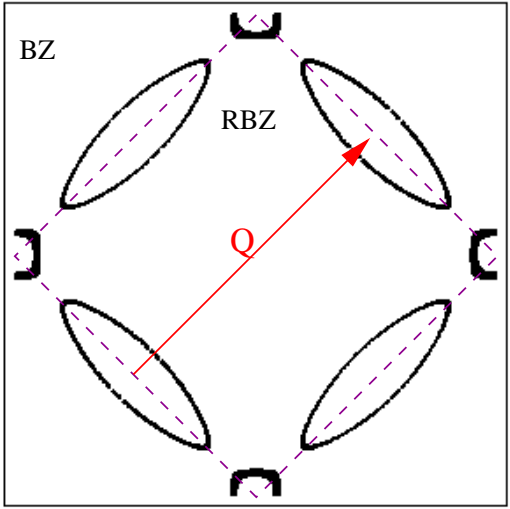}
\caption {The contour of constant energy for $\omega = 0.3D_0$ in the DDW model 
with $\mu=-t$, $D_0 = 0.1 t$ and $t' = -0.3t$.  
The dashed line encloses the reduced Brillouin zone (RBZ) and $\bQ$ is the DDW 
vector.  Note the electron pockets in the vicinity of $(\pm \pi,0)$ and $(0,\pm\pi)$.}
\label{fig:DDWBZ}
\end{figure}
%

%------------------------------------------------------------------------
\subsection{Competing orders}

As an example of a  model with competing orders we choose the $d$-density 
wave 
(DDW) theory.\cite{ddw1}  In this model, an order parameter of circulating
currents with $d$-wave symmetry competes with superconductivity and dominates 
the pseudogap phase.  The order parameter is in the particle-hole channel 
and connects excitations with momentum difference of $\bQ = (\pi,\pi)$ 
(see Fig.~(\ref{fig:DDWBZ})).  
The Green's function is conveniently written in the basis of 
$\psi^\dagger_\bk = (c^\dagger_\bk,c^\dagger_{\bk+{\bf Q}})$,
\begin{equation}\label{eq:DDWG}
G^0(\bk,i\omega)=[(i\omega-\epsilon'_\bk)-\epsilon''_\bk
\sigma_3 -D_\bk\sigma_2]^{-1},
\end{equation}
 with 
$\epsilon'_\bk={1\over 2}(\epsilon_\bk+\epsilon_{\bk+\bQ})$,
$\epsilon''_\bk={1\over 2}(\epsilon_\bk-\epsilon_{\bk+\bQ})$,
and the DDW gap
$D_\bk = {1 \over 2}D_0(\cos k_x -\cos k_y )$.
In this model, the Born approximation for the density of states modulations 
is given by
\begin{eqnarray}\label{eq:lamDDW}
\nonumber
\Lambda(\bq,i\omega) & =&
\tr {\sum_\bk}' [G^0(\bk,i\omega)(1+\sigma_1)
G^0(\bk-\bq,i\omega)]\\
&=&{2\over L^2}{\sum_\bk}' 
{-\Omega_+\Omega_-+\epsilon''_+\epsilon''_-+D_+D_-\over
(\Omega_+^2+E_+^2)(\Omega_-^2+E_-^2)},
\end{eqnarray}
where $i\Omega_\bk = i\omega-\epsilon'_\bk$ and the prime on the summation
indicates restriction to the reduced Brillouin zone depicted in Fig.~(\ref{fig:DDWBZ}). Derivation of this simple form is given in Appendix 
\ref{ap:DDW}. It is also worth noting that Eq.~(\ref{eq:lamDDW}), also used
by us previously, \cite{pbf} agrees with that used by Bena {\em et al.}\
\cite{bena1} in the appropriate limit of Born scattering.
\begin{figure}
\includegraphics[width = 8.5cm]{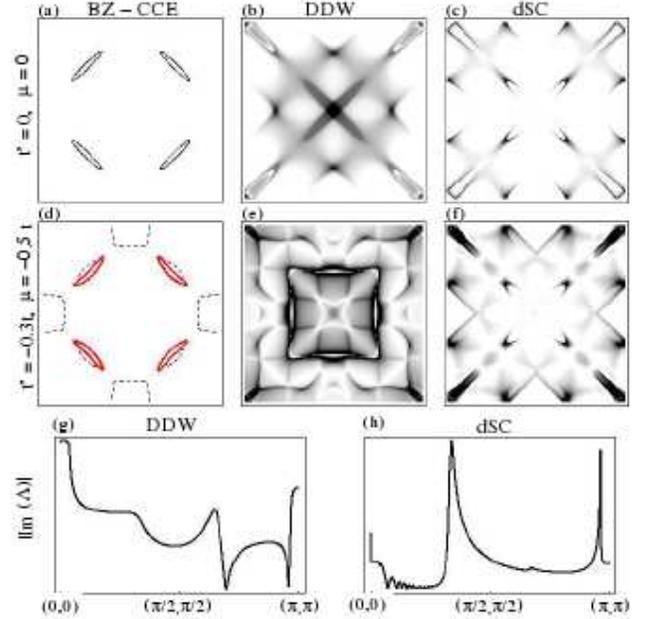}
\caption{ Interference patterns in a DDW model of the pseudogap compared 
with a $d$SC model. Panels (a-c,g,h) show the results in the limit $\mu = 0$ 
and $t' = 0$. Panel (a) represents the contour of constant energy at 
$\omega = 0.5\Delta_0$.  In this limit both models have the same CCE. Panels 
(b) and (c) show the calculated LDOS modulations in both models. These FT 
LDOS modulations differ from each other due to the different 
coherence factors as described in the text. These differences can also be 
seen in the one 
dimensional cut of the Brillouin zone along the $(\pi,\pi)$ direction, 
shown in panels (g) and (h).
Panel (d) shows the CCEs for the two different models when $t' = -0.3t$ and 
$\mu = -0.5t$.  In this case the energy dispersions are different; the 
solid line is the banana shaped CCE of the $d$SC model while the dashed line 
represents the relevant CCE of the DDW model. The calculated FT-STS for 
this case are presented in panels (e) and (f).
In all figures $\omega = 0.5\Delta_0$ and $\Delta_0=D_0 = 0.5t$. 
}
\label{fig:DDW}
\end{figure}

When the DDW gap is close to the Fermi energy, i.e., close to half filling, 
the low energy excitations exist only near the nodes of the DDW gap.  In this 
case we can linearize the spectrum in the vicinity 
of the nodes and approximate: 
$\epsilon'_\bk \to 0$, $\Omega_\bk \to \omega$, $\epsilon''_\bk\to v_F k_1$
and $D_\bk \to v_\Delta k_2$, where $k_1$ and $k_2$ are defined in the same 
way as in the $d$SC state.  The linearized form of Eq.~(\ref{eq:lamDDW}) turns
out to be  the same as that of a {\em magnetic} scatterer in the $d$SC state 
given by Eq.~(\ref{eq:lam5}).  This implies that close to half filling, a 
charge impurity in the DDW model is similar to a magnetic impurity in the 
$d$SC.
This behavior is due to charge conservation in the DDW model, i.e., there is 
no BCS mixing to produce the angular dependent coherence factors of the $d$SC 
state.

The expression (\ref{eq:lamDDW}) can also be evaluated exactly numerically 
for 
all dopings, and we have done so with the same band structure as in the 
previous cases.  The results are presented in Fig.~(\ref{fig:DDW})
along with the results for a $d$SC with the same parameters.

The analytical and numerical calculations of the LDOS modulations in the DDW 
model show two principal differences between this theory and the BCS theory 
of the $d$SC state.  First, the very different coherence 
factors lead to arc like patterns rather then peaks in the Brillouin zone. 
This feature cannot be altered by inclusion of higher order terms in the 
Born expansion (the $T$-matrix) due to the 
conservation of charge.  The solution for the $T$-matrix equation for the 
potential $V(1+\sigma_1)$ gives 
\begin{equation}
\hat{T}(\omega)=(1+\sigma_1){V\over 1-g_0(\omega) V}
\end{equation}
where $g_0(\omega) = {\rm Tr}\sum_\bk G^0(\bk,\omega)$. This does not 
affect the matrix structure of the scattering vertex but merely adds some 
frequency dependence to 
it. The second difference is in the positions of features in the Brillouin 
zone.  The positions are determined by the single particle dispersion, which 
in the DDW model is similar to that of a $d$SC state {\em only close to 
half filling} and in the absence of second nearest neighbor hopping $t'$.
Upon choosing more realistic band parameters the CCE of this model develop 
electron pockets and deviate from those of the $d$SC state, as shown in 
Fig.~(\ref{fig:DDW}-d).  This is a result of the fact that the node of the
DDW order parameter is {\em not} attached to the Fermi surface but to the 
$({\pi\over 2},{\pi\over 2})$ point in the Brillouin zone.

The patterns are most similar in the special case of $\mu=t'=0$, i.e.
exactly at half filling and with nearest neighbor hopping only. In this case
the excitation spectra for DDW and $d$SC are identical. The coherence factors,
however, are still different and give rise to qualitative differences
in the FT-STS patterns. As shown in Fig.~(\ref{fig:DDW}) away from this 
unphysical limit the patterns diverge
even more and it should be straightforward to distinguish between them 
experimentally.  This conclusion is independent of whether the 
superconducting state has coexisting DDW order or not.  This is due to the 
fact that the FT LDOS 
spectrum in the mixed state is very similar to that of a pure $d$SC as was 
shown by Bena {\it et al.}\cite{bena1}  

Our conclusions regarding the pure DDW state generally agree with those
of Ref.\ \onlinecite{bena1}, based on more detailed considerations within
the $T$-matrix approach. In particular our original finding,\cite{pbf} 
that the FT-STS patterns in DDW and $d$SC states are qualitatively 
different, seems to emerge independently of details of the band structure 
and the scattering mechanism. We argued above that such differences are even
more generic and will allow for discrimination between p-h and pairing
origin of the underlying electron order in the pseudogap state.  

%-------------------------------------------------------------------------
\subsection{Relation to experimental data}

Recently, two groups succeeded in obtaining the FT-STS data in the pseudogap 
state of two different cuprate superconductors. Vershinin 
{\em et al.}\cite{vershinin1} focused on 
the $T>T_c$ region in weakly underdoped samples of 
Bi$_2$Sr$_2$CaCu$_2$O$_{8+\delta}$ (BiSCCO), while Hanaguri {\em et al.}
\cite{hanaguri1} studied lightly doped (non-superconducting) single crystals
of Ca$_{2-x}$Na$_x$CuO$_2$Cl$_2$ (Na-CCOC) at low temperatures. In both
cases no dispersive features could be identified.
Instead, non-dispersive patterns with periodicity close
to 4 lattice spacings have been reported in both experiments and interpreted 
as evidence for underlying static charge ordering. Precise nature of this 
ordering phenomenon is unclear at present and two scenarios involving a 
hole Wigner crystal\cite{fu1} and a pair density wave\cite{chen1,tesanovic1}
have been proposed.

Absence of the quasiparticle interference patterns in Refs.
\onlinecite{vershinin1,hanaguri1} implies that the methods described in 
this work are not directly applicable to these experiments. 
A more promissing in this respect is the recent experiment of McElroy  
{\em et al.}\cite{mcelroy1}, which identifies `zero temperature pseudogap'
spectra in nominally superconducting samples of BiSCCO an low temperatures.
Such ZTPG spectra coexist with the more conventional superconducting spectra
and form characteristic nanoscale patchwork of interconnected regions. 
In strongly 
underdoped BiSCCO regions exhibiting ZTPG almost completely 
dominate the field of view. The interesting aspect of this data is that 
at low energies these ZTPG regions exhibit FT-STS interference patterns 
that look just like those observed deep in the superconducting state. At
higher energy, however, non-dispersive peaks with periodicity close
to 4 lattice spacing are observed in these same regions. These observations
indicate that (i) the low energy quasiparticles in the ZTPG regions are of
superconducting Nambu-Gorkov variety, and (ii) there exists an
intimate relationship between the superconducting order and 
the new ordering phenomenon these materials.

An interesting question arises as to why are the dispersive peaks absent 
in the patterns observed in  
the pseudogap state. One possibility is that there are simply no sufficiently 
coherent quasiparticles that could form the interference patterns (such as
in the QED$_3$ theory with $\eta>{1\over 2})$. Another possibility is that 
the patterns are present but have not been so far detected. It 
is entirely possible that in the BiSCCO experiment\cite{vershinin1} done at 
$T\sim 100$K delicate interference patterns are 
washed out by thermal broadening. The Na-CCOC experiment,\cite{hanaguri1}
on the other hand, is done at $T\sim 0.1$K and thermal smearing 
cannot be invoked.
In this case it is interesting to note that no dispersive patterns have 
been observed in this material even in the superconducting state
at higher doping. These considerations thus raise hopes that the interference
patterns may be observed in future experiments and could help determine the 
nature of the pseudogap state in cuprates.

%%%%%%%%%%%%%%%%%%%%%%%%%%%%%%%%%%%%%%%%%%%%%%%%%%%%%%%%%%%%%%%%%%%%%%%%%%%
\section{Conclusions}

We have shown that the quasiparticle interference 
patterns seen in the FT-STS experiments reveal signatures of both the 
quasiparticle dispersion and, through their sensitivity to
the quasiparticle quantum phases, the {\em nature of the underlying 
electronic order} in the system. 
In particular, the quasiparticle phase determines the coherence factors, 
which then determine the basic characteristic features of the interference 
patterns. We have demonstrated, by general arguments and
detailed calculations within two relevant models, that the 
superconducting order alone produces patterns consistent with the existing
experimental data. In particular, order of another type (such as DDW) yields
{\em qualitatively different} patterns, even if the quasiparticle excitation 
spectrum is identical to the $d$SC.

The above feature is of key importance since it is in general difficult 
to model the details of FT-STS patterns, even in the well understood $d$SC
state. This difficulty presumably arises from the complexities of the 
scattering processes which in realistic materials are likely caused by
several distinct mechanisms and must be modeled as a combination of
magnetic, non-magnetic and pairing potentials. In addition the matrix elements
for the tunneling between the sample and the STS tip, ignored in most 
theoretical treatments, are likely to affect the patterns. 

Despite these complexities, the FT-STS carries a wealth of useful information.
Sensitivity of the 
interference patterns to the type of ordering implies that even though we 
might not be able to model their every detail in a particular state, we are 
assured that {\em the patterns will change} as the material crosses
a phase boundary to a state characterized by a different electronic order.

Based on this
insight we have proposed a test for the nature of the pseudogap phase
in cuprates using FT-STS. The phase boundary in this case is 
$T_c(x)$, i.e.\ the superconducting critical temperature as a function 
of hole doping $x$. If the pseudogap is due to fluctuating SC 
order, then the FT-STS patterns above $T_c$ should remain 
qualitatively the same as those below $T_c$. If, on 
the other hand, the pseudogap is due to an order parameter in the
particle-hole channel, such as SDW, CDW or DDW, the patterns above $T_c$ should
be {\em qualitatively different}, due to their different coherence factors.

Although we focused here on one example of a particle-hole order parameter, 
the arguments presented above are general and can be applied to distinguish 
any such order from superconductivity. In addition, DDW order is perhaps 
the most germane to our argument since for specially selected parameters 
the excitation spectrum is the same as in the $d$SC. We have demonstrated that
even in this special case the FT-STS patterns exhibit qualitative differences 
and should be easily distinguishable from the $d$SC. This is a result of the 
quasiparticle charge 
non-conservation in the superconductor which manifests itself in the
coherence factors of the FT-STS  spectra. Other types of p-h order, such as 
SDW and CDW, will generally have very different excitation spectra, 
trivially implying different interference patterns.

By probing the interference patterns below and above $T_c(x)$, future
FT-STS experiments should be able to unambiguously discriminate between 
the remnants of the superconducting order and an order in the particle-hole
channel, thus settling one of the key puzzles in the high-$T_c$ 
superconductivity.

%----------
\smallskip
\noindent
{\it Acknowledgments\/} ---  
%---------
The authors are indebted to J.C. Davis, T.P. Davis, V.P. Gusynin, J. 
Hoffman, A. Iyengar, D.-H. Lee, 
S. Sachdev, D.E. Sheehy, O. Vafek, Z. Te\v{s}anovi\'c and A. Yazdani
for discussions and correspondence. This work was supported by NSERC,
CIAR, the A.P. Sloan Foundation and partly by the NSF under grant No. 
PHY99-07949 (preprint NSF-KITP-04-11). One of the authors (M.F.) wishes to 
acknowledge the hospitality 
of the Aspen Center for Physics, where part of the work was performed and
many stimulating discussions took place.

%%%%%%%%%%%%%%%%%%%%%%%%%%%%%%%%%%%%%%%%%%%%%%%%%%%%%%%%%%%%%%%%%%%%%%%%%%
\appendix
\section{Linearized approximation}\label{inter-nodal}

\subsection{Non-magnetic scattering}

Let us first consider the density of states modulations resulting from 
scattering within the same node in the linearized approximation given in 
Eq.~(\ref{eq:lin7}).  The solution, Eq.~(\ref{eq:lin7s}), can be found by 
``combining the denominators'' using the Feynman 
parameterization.\cite{peskin} This is based on the formula
\begin{equation}
{1 \over A^\alpha B^\beta} = {\Gamma(\alpha +\beta) \over \Gamma(\alpha)\Gamma(\beta)}
\int_0^1 dx {x^{\alpha-1}(1-x)^{\beta-1} \over [xA+(1-x)B]^{\alpha+\beta}},
\end{equation}
where $\Gamma(z)$ is the Gamma function and the identity is valid for any positive $A, B, \alpha, \beta$.
Applying this to Eq.~(\ref{eq:lin7}) with $\alpha=\beta=1$ and performing a 
shift in the integration variable  $\bk \to \bk +(1-2x)\tbq $ yields
%wwwwwwwwwwwwwwwwwwwwwwwwwwwwwwwwwwwwwwwwwwwwww
\begin{widetext}
\begin{equation}
\Lambda_{\rm lin}(\bq,i\omega) = {1\over v_Fv_\Delta} \int_0^1 dx  
\int {d^2k\over(2\pi)^2}
{-\omega^2 + (k_1^2-k_2^2)-4x(1-x)(\tq_1^2-\tq_2^2)\over
[\omega^2+ \bk^2+4x(1-x)\tbq^2]^2}.
\end{equation}
The $k_1^2-k_2^2$ term vanishes in the angular integration and the radial 
integral is convergent.  
After performing the integration over momentum we get
\begin{equation}
\Lambda_{\rm lin}(\bq,i\omega) =
{1\over 4\pi v_Fv_\Delta}\int_0^1 dx 
{-\omega^2-4x(1-x)(\tq_1^2-\tq_2^2)\over
\omega^2+4x(1-x)\tbq^2}
\end{equation}
and the solution quoted in the text follows by performing the $x$ integration 
and Wick rotating the frequency $i\omega \to \omega$ to get the retarded 
quantity. 

Now consider a scattering process between two antipodal nodes, spanned by the 
vector $\bQ$ (shown in Fig.~(\ref{fig:nodes})) connecting the nodes along the 
$(\pi,\pi)$ direction (the length of $\bQ$ varies with the doping).  
The linearized form of $\Lambda$ is similar to Eq.~(\ref{eq:lin7s}) with some 
signs inverted
due to the different directions of $v_F$ and $v_\Delta$ near the two nodes
\begin{eqnarray}\label{eq:lin34}
\nonumber
\Lambda_{\rm lin}(\bq+\bQ,i\omega) &=& {1\over v_Fv_\Delta}
\int {d^2 k\over(2\pi)^2} 
{-\omega^2 - (k_1^2-k_2^2)+(\tq_1^2-\tq_2^2)\over
[\omega^2+(\bk-\tbq)^2][\omega^2+(\bk+\tbq)^2]} \nonumber \\
&=&{1\over 4\pi v_Fv_\Delta}\int_0^1 dx 
{-\omega^2+4x(1-x)(\tq_1^2-\tq_2^2)\over
\omega^2+4x(1-x)\tbq^2},
\end{eqnarray}
where in the last line we have introduced the Feynman parameter $x$ and 
performed the momentum integral. The result is given by Eq.~(\ref{eq:lin7s}) 
with the angular factor replaced by 
$\tilde q_1 ^2 / \tilde q^2$.  This $90^{\rm o}$ rotation of the angular factor 
causes the high intensity regions
to be two points on the elliptic contour in the 
direction parallel to the vector $\bQ$ as shown in Fig.~(\ref{fig:lin}b).  
This is a deviation with respect to the octet model. Instead of the vectors 
$\bq_3$ and $\bq_4$ connecting the edges of the antipodal CCE bananas we 
obtain $\tbq_3 \approx \bq_3$ at the points 
$\bQ \pm {2\omega\over v_F}\hat q_1$.  The octet vector $\bq_4$ is close to 
the point where the angular coherence factor vanishes and hence appears as 
an end of a line rather than a peak. 

In the two cases above we have scaled the momenta $k_1$ and $k_2$ by $v_F$ 
and $v_\Delta$ respectively 
to achieve radial symmetry. This convenient scaling can not be applied to 
the case of adjacent node 
scattering. For generic anisotropy one then obtains more complicated
expressions which cannot be evaluated in a closed form.   To keep things
simple we treat this process in the limit of isotropic velocity 
$v_F = v_\Delta \equiv v$. If we denote by $\bQ'$ the 
vector  that connects two adjacent nodes we have 
\begin{equation}\label{eq:lin12}
\Lambda_{\rm lin}(\bq+\bQ',i\omega) = {1\over v^2}\int {d^2 k\over(2\pi)^2} 
{-\omega^2 \over [\omega^2+(\bk-\tbq)^2][\omega^2+(\bk+\tbq)^2]},
\end{equation}
\end{widetext}
%eeeeeeeeeeeeeeeeeeeeeeeeeeeeeeeeeeeeeeeeeeeeeeee
where we have omitted terms that are odd functions of $\omega$. These would 
cancel later, when taking the appropriate components of $\Lambda$
according to the prescription in Eq.~(\ref{eq:components}).
The solution is given by
\begin{eqnarray}\label{eq:lin12s}
\Lambda_{\rm lin}(\bq,\omega) &=&{1\over 4\pi v^2}
\left({\omega \over\tq }\right)^2
{\cal S}\left({\omega\over\tq}\right)
,\nonumber\\
{\cal S}(z)&=&{1\over\sqrt{z^2-1}}\arctan{1\over\sqrt{z^2-1}}.
\end{eqnarray}
There is no angular dependence and the resulting $\Lambda$ is just a 
uniform contour, as shown in Fig.~(\ref{fig:lin}d).
More detailed considerations indicate that restoring the anisotropy would 
introduce a four-fold modulation of the intensity along the contour.

%------------------------------------------------------------------------------
\subsection{Magnetic scattering}

The inter-nodal magnetic scattering can be calculated in a similar way for a 
magnetic point like impurity, where the only difference
is the sign in front of the $\Delta_+\Delta_-$ term in Eq.~(\ref{eq:lam2}).  
This leads to the linearized form in Eq.~(\ref{eq:lam5}) from which it can be 
readily seen that no angular dependent coherence factors emerge in this case.
  The solution, Eq.~(\ref{eq:lam5s}) is obtained by following the same steps 
as in the previous subsection.
For antipodal nodes the result is the same as for the intra-nodal magnetic 
scattering, Eq.~(\ref{eq:lam2}), where there is no angular dependence.  
However, when considering the case of scattering between adjacent nodes, the 
magnetic scattering 
exhibits angular dependence that generates peaks close to the octet model 
vectors $\bq_1,\bq_5$ and  $\bq_{2,6}$ even in the isotropic limit.
The solution for this case is given by: 
\begin{eqnarray}\label{eq:lin7m}
\Lambda_{\rm lin}^{\rm mag}(\bq,\omega) &=&{1\over 4\pi v^2}
\left({\tq_1\tq_2+\omega^2 \over\tq^2}\right) 
{\cal S}\left({\omega\over\tq}\right)
,
\end{eqnarray}
where ${\cal S}(z)$ is defined in Eq.~(\ref{eq:lin12s}) and the factor of 
$\tq_1\tq_2/\tq^2$ 
is responsible for four peaks along the contour, near the octet vectors $\bq_1,\bq_5$, and $\bq_{2,6}$ as shown in Fig.~(\ref{fig:lin}c).

The above analysis suggests that a combination of non-magnetic and magnetic
scattering is needed to produce sharp peaks at all the octet 
vectors. The exception is $\bq_4$ which marks the end of a line produced 
by Eq.~(\ref{eq:lin34}). The linearized approximation results are given in 
Fig.~(\ref{fig:lin}). In panel (e) we have also combined together the 
results for non-magnetic scattering to cover the entire Brillouin zone. 
For adjacent node scattering we use magnetic results which show 
the four-fold modulation expected for the anisotropic case.

%--------------------------------------------------------------
\subsection{QED$_3$}

The nodal approximation is also useful for the case of propagators with 
anomalous dimensions, as in the case of QED$_3$.  Let us consider the 
intra-nodal scattering processes within the QED$_3$ framework.
We use the same Feynman parameterization as before, where this time 
$\alpha = \beta = 1-\eta/2$. This leads to the following integral
%wwwwwwwwwwwwwwwwwwwwwwwwwwwwwwwwwwwwwwwwwwwwwwwwww
\begin{widetext}
\begin{eqnarray}
\nonumber
\Lambda_{\eta}(\bq,i\omega) &=& {1\over v_F v_\Delta }\int 
{ d^2 k \over (2\pi)^2} 
{(i\omega+k_1+\tq_1)(i\omega+k_1-\tq_1) 
\over [\omega^2+(\bk+\tbq)^2]^{1-\eta/2}[\omega^2+(\bk-\tbq)^2]^{1-\eta/2}} \\ 
&=& {1\over v_F v_\Delta }{\Gamma(2-\eta)\over (\Gamma(1-\eta/2))^2}
\int _0^1 dx [x(1-x)]^{-\eta/2} 
\int { d^2 k \over (2\pi)^2} {-\omega^2+k_1^2-4x(1-x)\tq_1^2 
\over [\omega^2+\bk^2 +4x(1-x)\tbq^2]^{2-\eta}}
\end{eqnarray}
where the last line was obtained by the variable shift $
\bk \to \bk - (2x-1)\tbq$.
We can now evaluate the momentum integral by first performing the angular
 integration and then the radial part with a momentum cutoff $\lambda$.  
We obtain
\begin{eqnarray}
\int { d^2 k \over (2\pi)^2} \dots
&=& {1\over 4\pi (1-\eta)}\left( {\omega^2+4x(1-x)\tq_1^2 \over 
[\omega^2+4x(1-x)\tbq^2]^{1-\eta}} + 
{1\over 2\eta}[\lambda^{2\eta} - (\omega^2+4x(1-x)\tbq^2)^\eta]\right) \\ 
&=& {1\over 4\pi (1-\eta)}\left([\omega^2+4x(1-x)\tbq^2]^\eta -
{4x(1-x)\tq_2^2 \over [\omega^2+4x(1-x)\tbq^2]^{1-\eta}} + 
{1\over 2\eta}[\lambda^{2\eta} - (\omega^2-4x(1-x)\tbq^2)^\eta]\right), 
\nonumber
\end{eqnarray}
where only the middle term is divergent.  
We therefore proceed considering the divergent term only,
\begin{equation}
\Lambda_\eta(\bq,i\omega)|_{div} = 
{1\over (1-\eta)}{1\over 4\pi v_F v_\Delta }{\Gamma(2-\eta)\over (\Gamma(1-\eta/2))^2}
\int _0^1 dx [x(1-x)]^{-\eta/2} 
{4x(1-x)\tq_2^2 \over [\omega^2+4x(1-x)\tbq^2]^{1-\eta}}.
\end{equation}
\end{widetext}
%eeeeeeeeeeeeeeeeeeeeeeeeeeeeeeeeeeeeeeeeeeeeeeeeeeeeeeeeeee
By performing the Wick rotation and using $z^2 = \omega^2/\tbq^2$ we 
find
\begin{equation}\label{eq:lambda-eta}
\Lambda_\eta(\bq,\omega)\propto 
\left({\tq_2 \over \tbq}\right)^2 \int _0^1 dx {[x (1-x)]^{1-\eta/2} 
\over [x(1-x)-({z\over2})^2]^{1-\eta}}.
\end{equation}
The angular dependence of $\Lambda$ along the elliptic contour is already 
explicit in the above expression and is the same as in the $d$SC intra-nodal 
scattering case.  Unfortunately, the Feynman parameter integration can not
be performed analytically for all values of $z$.  For the case of $z<1$ 
Eq.~\ref{eq:lambda-eta} can be evaluated using hypergeometric functions 
but no such expansion exists for $z\geq 1$. We therefore choose to analyze the 
integral and find its degree of divergence.  
First note that the denominator is divergent 
whenever $z^2 =4 x(1-x)$. For any given value of $z^2$ this leads to two 
singularities at $x/2 = -1 \pm \sqrt{1-z^2}$. However, since $\eta$ is 
positive and smaller than 1, the singularities are integrable unless 
they overlap.  Therefore, the integral is only divergent when $z^2 = 1$.  
In order to find the degree of divergence we shift the integration variable 
$x \to x-1/2$ and denote $b = (1-z^2)/4$. The $x$ integral, $I(z)$, is given 
by
\begin{eqnarray}
I(z) &=& 2 \int _0^{1\over 2} dx {({1\over 4}-x^2)^{1-\eta/2} 
\over (b-x^2)^{1-\eta}} \\
&=&2 \left({1 \over b}\right)^{1-\eta}\int_0^{1\over 2} dx 
{({1\over 4}-x^2)^{1-\eta/2} 
\over [1-({x\over \sqrt b})^2]^{1-\eta}}.
\nonumber 
\end{eqnarray}
We can now define a new integration variable $y={x\over \sqrt b}$ and write
\begin{eqnarray}
\nonumber 
I(z)  
=2 \left({1 \over b}\right)^{{1\over 2}-\eta}\int_0^{1\over2\sqrt b}dy 
{({1\over 4}-by^2)^{1-\eta/2} \over (1-y^2)^{1-\eta}}.
\end{eqnarray}
The remaining integral is convergent in the limit $b\to 0$. 
The integration thus can be performed on $[0,\infty)$ and the singularity, 
given by the prefactor, is of the order 
of $1/(1-z^2)^{1/2 - \eta}$, as stated in the text.

%-----------------------------------------------------------------------
\section{Gap suppression in a $d$-wave superconductor}
\label{bond}

\begin{figure}
\includegraphics[width = 6.5cm]{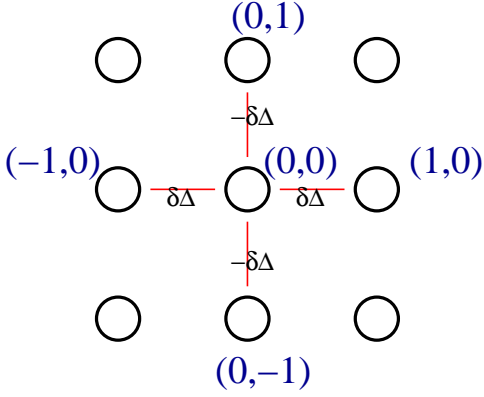}
\caption{The gap modulation local perturbation around the lattice point 
$\bR=(0,0)$.}\label{fig:gapmod}
\end{figure}

In order to take into account the gap modulations in the sample we introduce 
a local perturbation which will 
be referred to as the {\it gap suppression}.   
We start with a lattice model where both the hopping amplitudes 
$t$, $t'$ and the gap function $\Delta_{ij}$ are defined on the bonds.
We assume that $\Delta_{ij} = \pm \Delta_0$ on nearest neighbors and is zero
 everywhere else, with plus and minus sign corresponding to $x$ and $y$ oriented bonds 
respectively.  On top of this uniform state we introduce a gap suppression  
$\delta\Delta(\bR)$ on four bonds around the point $\bR$, 
as shown in Fig.~(\ref{fig:gapmod}).  
The distribution of these impurities is assumed to be a featureless function 
of $\bR$. The gap modulation produces the following perturbation: 
\begin{eqnarray}
\nonumber
\delta{\cal H} &=& \sum_{\langle\bR,\bR'\rangle}[\delta\Delta(\bR)c^\dagger_{\bR}
c^\dagger_{\bR'}\chi(\bR-\bR')+\rm{h.c.}]
 \\ &=& \sum_{\bp,\bk}[\delta\Delta_\bp\chi_\bk c^\dagger_\bp c^\dagger_\bk +
\rm{h.c.}],
\end{eqnarray}
where $\langle..,..\rangle$ denotes the  nearest neighbors on the lattice, 
$\chi(\bR-\bR')$ is the sign of the 
perturbation, depending on the bond orientation and 
$\chi_\bk = \cos k_x -\cos k_y $.  
To leading order, this potential leads to the following LDOS modulation
\begin{eqnarray}\label{eq:components2}
\delta n(\bq,\omega) =-{1\over \pi} |\delta\Delta_\bq|  
{\rm Im} \left[\Lambda_{11}(\bq,\omega)+ \Lambda_{22}(\bq,-\omega) \right],
\end{eqnarray}
with
\begin{equation}
\Lambda(\bq,\omega)=\sum_\bk G^0(\bk,\omega)\chi_\bk\sigma_1
G^0(\bk-\bq,\omega).
\end{equation}
Note that the perturbation $\chi_\bk\sigma_1$ is only a function of one 
momentum index, conjugate to the relative coordinate $\bR-\bR'$.

One can go further and find the $T$-matrix for a combined local potential of a 
charge impurity and a gap suppression. For simplicity, we assume that the same 
dilute array of impurities is responsible for both perturbations and treat the
 single impurity at the origin in the $T$-matrix approach.
For the combined potential $ V_{\bk\bk'} = V\sigma_3 + 
\delta\Delta\chi_\bk\sigma_1$ the $T$-matrix equation (\ref{t1}) 
can be solved using the ansatz
\begin{equation}
\hat{T}_{\bk\bk'}(\omega) = T_0(\omega)\sigma_0 + T_3(\omega)\sigma_3 +
T_1(\omega)\chi_\bk\sigma_1.
\end{equation}
The frequency dependent coefficients are given by:
\begin{eqnarray}
\nonumber
T_0(\omega)&=&{g_0(\omega) V^2 \over A(\omega)},\\
\nonumber 
T_3(\omega)&=&V{1-g_1(\omega) V \over A(\omega)},\\ 
\nonumber
T_1(\omega)&=&\delta\Delta{1-g_1(\omega) V \over A(\omega)},
\end{eqnarray}
with
\begin{eqnarray}
\nonumber
A(\omega) &\equiv& [1-g_1(\omega) V]^2-[g_0(\omega) V]^2 \\ &&-
g_d(\omega) \delta\Delta[1- g_1(\omega) V].
\end{eqnarray}
$g_i(\omega)$ are functions of the frequency, resulting from the
momentum integration of different parts of the unperturbed Green's function:
\begin{eqnarray}
\nonumber
g_0(\omega) &=& \sum_\bk {\omega \over \omega^2 - E_\bk^2}\\
\nonumber
g_1(\omega) &=& \sum_\bk {\epsilon_\bk \over \omega^2 - E_\bk^2}\\
g_d(\omega) &=& \sum_\bk {\chi_\bk\Delta_\bk \over \omega^2 - E_\bk^2}.
\end{eqnarray}
This $T$-matrix has a resonant structure in frequency, similar to that found 
by Balatsky {\it et al.} for the case of a charge impurity.\cite{Balatsky1}
However, we are interested in a particular regime of frequencies and 
scattering strength which may not coincide with the resonance.  
The advantage of this formalism in this case is the mixture of different 
scattering components and the finite imaginary part of the $T$-matrix.

%--------------------------------------------------------------------------

\section{Quasiparticle scattering in d-Density wave model}\label{ap:DDW}
In order to calculate $\Lambda(\bq,\omega)$ in the DDW model one has to 
carefully define quantities in 
the reduced Brillouin zone (RBZ) shown in Fig.~(\ref{fig:DDWBZ}).
Our starting point will be the unperturbed DDW Hamiltonian\cite{ddw1}
\begin{eqnarray}\label{eq:DDWH}
\nonumber
{\cal H}& = &\sum_\bk [ \epsilon_\bk^{} a^{\dagger}_\bk a^{}_\bk +iD_\bk^{} 
a^{\dagger}_\bk a^{}_{\bf k+Q}
-iD_\bk^{} a^{\dagger}_{\bf k+Q}a^{}_\bk] \\
&\equiv &{\sum_\bk}' \psi^{\dagger}_\bk{\cal H}^{}_\bk\psi^{}_\bk
\end{eqnarray}
where $a_\bk$ are the electron operators, the prime denotes a sum over RBZ 
wave vectors and $D_\bk$ is the DDW order parameter.  We have defined 
$\psi^{\dagger}_\bk = (c^{\dagger}_\bk,d^{\dagger}_\bk)$ such 
that $c_\bk$ is the usual electron annihilation operator, $a_\bk$ when 
$\bk$ is in the reduced 
Brillouin zone and $d_{\bf k+Q}$ is the usual electron annihilation operator 
when $\bk$ is outside of the RBZ (and ${\bf k+Q} \in$ RBZ).  The operators 
$c$ and $d$ 
are defined only in the RBZ and therefore their labels should be understood 
modulo ${\bf Q} = (\pi,\pi)$. In what follows we maintain the labels of $c$, 
$d$ and $\psi$
explicitly in the RBZ.  In this basis the matrix ${\cal H}_\bk$ is given by:
\begin{equation}
{\cal H}_\bk = 
\begin{pmatrix}
        \epsilon_\bk\,~~~-iD_\bk\\ iD_\bk\, ~~~\epsilon_{\bf k+Q}
\end{pmatrix} 
\end{equation}
For simplicity we consider point scattering described by the Hamiltonian  
\begin{eqnarray}
\delta{\cal H} &=& \sum_\br V\delta(\br) a^\dagger_\br a_\br^{} =
V\sum_{\bp,\bp'}a^\dagger_\bp a_{\bp'}^{}\\ \nonumber
&=&V{\sum_{\bp,\bp'}}'\psi^{\dagger}_\bp(1+\sigma_1)\psi^{}_{\bp'}.
\end{eqnarray}
The same conclusion can be derived for arbitrary weak potential.

In order to calculate the tunneling amplitude we  use the interaction 
representation, where the DDW Hamiltonian $\cal H$ represents the uniform 
system and $\delta{\cal H}$ is the interaction.  We express the Green's 
function in terms of RBZ momenta and expand it to first order in the 
interaction. The single particle Green's function is given by
\begin{equation}
G(\br,\br';t-t')=-i\frac{\langle T a^{}_\br(t) a^\dagger_{\br'}(t')S(\infty,-\infty) \rangle}
 {\langle S(\infty,-\infty) \rangle}
\end{equation}
where the $S$-matrix is responsible for the time evolution of operators by 
the interaction $\delta{\cal H}$, the angular brackets denote the vacuum 
expectation values and $T$ is the time ordering operator.  We shall now 
focus on the equal point propagator $G(\br,\br;t-t')$ and consider its
Fourier transform with respect to $\br$
%wwwwwwwwwwwwwwwwwwwwwwwwwwwwwww
\begin{widetext}
\begin{equation}
\tilde G(\bq;t-t')\equiv \int d \br e^{-i \bq \cdot \br} G(\br,\br;t-t')
= {-i \over \langle S(\infty,-\infty) \rangle} 
\sum_\bk \langle T a^{}_\bk(t) a^\dagger_{\bk-\bq} (t')S(\infty,-\infty) \rangle.
\end{equation}
The momentum indices are defined in the full Brillouin zone, where one of 
the following four cases may occur. Both $\bk$ and $\bk-\bq$ are in the RBZ, 
both are outside of the RBZ, $\bk \in$ RBZ and $\bk-\bq \notin$ RBZ or 
$\bk \notin RBZ$ and $\bk-\bq \in RBZ$.  In the RBZ we can write this as
\begin{eqnarray}\label{ddw1}
\nonumber
\tilde G(\bq;t-t')& =& {-i \over \langle S(\infty,-\infty) \rangle} 
{\sum_\bk}'  \langle T\{ [c^{}_\bk(t) c^\dagger_{\bk-\bq}(t')+d^{}_\bk(t) d^\dagger_{\bk-\bq}(t')]\Theta(\bk-\bq) +   \\ 
&&\;\;\;\;\;[c^{}_\bk(t)d^\dagger_{\bk-\bq+\bQ}(t')+d^{}_\bk(t)c^\dagger_{\bk-\bq+\bQ}(t')]\bar\Theta(\bk-\bq)\}S(\infty,-\infty) \rangle
\\
 &=&{-i \over \langle S(\infty,-\infty) \rangle} 
{\rm Tr}{\sum_\bk}'  \langle T\{ \psi^{}_\bk(t) \psi^\dagger_{\bk-\bq}(t')\Theta(\bk-\bq) + \sigma_1\psi^{}_\bk(t) \psi^\dagger_{\bk-\bq+\bQ}(t')\bar\Theta(\bk-\bq) \}(t')S(\infty,-\infty) \rangle.
\nonumber
\end{eqnarray}
where
\[
\Theta(\bk)=\biggl\{
\begin{array}{ll}
1, & \  \bk\in{\rm RBZ} \\
0, & \  \bk\notin{\rm RBZ} 
\end{array}, \ \ \ \ \bar\Theta(\bk)=1-\Theta(\bk).
\]
 In the last line of (\ref{ddw1}) we introduced the $\sigma_1$ Pauli matrix 
in the trace in order to implement the sum over off-diagonal elements of the 
$2\times2$ matrix $\psi_\bk^{}\psi^\dagger_{\bk-\bq}$.  

We now include the perturbation in the $S$-matrix to first order, i.e., 
we approximate 
$S(\infty,-\infty) \approx 1-i \int\delta {\cal H} dt$.  Dropping the 
unperturbed part of the Fourier transformed Green's function we find the 
first order correction
\begin{equation}  
\tilde G^{(1)}(\bq;t-t') =  -V{\rm Tr}{\sum_{\bk  \bp \bp'}}' \int dt''  \langle T\{
 \psi^{}_\bk(t) \psi^\dagger_{\bk-\bq}(t')\Theta(\bk-\bq) + \sigma_1\psi^{}_\bk(t) \psi^\dagger_{\bk-\bq+\bQ}(t')\bar\Theta(\bk-\bq)\} \psi^{\dagger}_\bp(t'')(1+\sigma_1)\psi^{}_{\bp'}(t'') \rangle.
\end{equation}
Performing Wick contractions and a time Fourier transform yields the quantity $\Lambda(\bq,\omega)$, defined in the text:
\begin{eqnarray}
\nonumber  
\Lambda(\bq,\omega) &=& {1\over V}\int dt \; e^{-i\omega (t-t')} G^{(1)}(\bq;t-t')   \\ \nonumber
&=&{\rm Tr}{\sum_\bk }' \int dt'' dt \; e^{-i\omega (t-t')} [ G^0(\bk,t-t'')(1+\sigma_1)G^0(\bk-\bq,t''-t')\Theta(\bk-\bq) \\ 
&+& \sigma_1G^0(\bk,t-t'')(1+\sigma_1)G^0(\bk-\bq+\bQ,t''-t')\bar\Theta(\bk-\bq)] \\
&=& {\rm Tr}{\sum_\bk }'  [ G^0(\bk,\omega)(1+\sigma_1)G^0(\bk-\bq,\omega)\Theta(\bk-\bq) + \sigma_1G^0(\bk,\omega) (1+\sigma_1)G^0(\bk-\bq+\bQ,\omega)\bar\Theta(\bk-\bq) ] 
\nonumber
\end{eqnarray}
where $G^0(\bk,\omega)$ is the unperturbed DDW Green's function 
given in Eq.~(\ref{eq:DDWG}). 
The above expression can be simplified further.  Let us 
concentrate on the second term when $\bk-\bq \notin$ RBZ.  If we now extend 
the definition of the Green's function to hold even when the argument is not 
in the RBZ (using the functional definitions of 
$\epsilon_\bk$ and $D_\bk$), we find that 
$G^0(\bk+\bQ,\omega) = \sigma_1 G^0(\bk,\omega) \sigma_1$.
Rewriting the second term we find:
\begin{eqnarray}
 \nonumber
\tr [\sigma_1 G^0(\bk,\omega)(1+\sigma_1)G^0(\bk+\bq+\bQ,\omega)]&=&
\tr [\sigma_1 G^0(\bk,\omega)(1+\sigma_1)\sigma_1 G^0(\bk+\bq,\omega)\sigma_1]\\
&=&\tr [G^0(\bk,\omega) (1+\sigma_1) G^0(\bk+\bq,\omega)].
\end{eqnarray}
Therefore
\begin{equation}
\Lambda(\bq,\omega)={\sum_\bk}'\tr [G^0(\bk,\omega) (1+\sigma_1) G^0(\bk+\bq,\omega)],
\end{equation}
where $\bk+\bq$ need not be in the RBZ.
\end{widetext}
%wwwwwwwwwwwwwwwwwwwwwwwwwwwwwwww


\begin{thebibliography}{10}
\bibitem{emery1} V.J. Emery and S.A. Kivelson, Nature {\bf 374},
434 (1995).
\bibitem{balents1} L. Balents, M.P.A. Fisher and C. Nayak, 
\prb {\bf 60}, 1654 (1999).
\bibitem{ft1} M.~Franz and Z.~Te\v{s}anovi\'c, \prl {\bf 87}, 257003 (2001).
\bibitem{so5} S.C. Zhang, Science {\bf 275}, 1089 (1997).
\bibitem{laughlin1} R. B. Laughlin, cond-mat/0209269.
\bibitem{ddw1} S. Chakravarty, R.B. Laughlin, D.K. Morr, and 
C. Nayak \prb {\bf 63}, 094503 (2001).
\bibitem{timusk1} T. Timusk and B.W. Statt,  Rep. Prog. Phys. {\bf 62},
61  (1999).
\bibitem{Sutherland} M. Sutherland, D. G. Hawthorn, R. W. Hill, F. Ronning, 
S. Wakimoto, H. Zhang, 
C. Proust, E. Boaknin, C. Lupien, L. Taillefer, R. Liang, D. A. Bonn, 
W. N. Hardy, R. Gagnon, N. E. Hussey, T. Kimura, M. Nohara, and H. Takagi
 \prb {\bf 67}, 174520 (2003).
\bibitem{Valla} T.~Valla, A.~V.~Fedorov, P.~D.~Johnson, Q.~Li, G.~D.~Gu, and 
N.~Koshizuka \prl {\bf 85}, 828 (2000).
\bibitem{varma1} C.M. Varma, \prl {\bf 83}, 3538 (1999).  
\bibitem{vojta1} M. Vojta, Y. Zhang, and S. Sachdev, \prb {\bf 62},
6721 (2000).
\bibitem{randeria1} M. Randeria, {\em Varenna Lectures} 
(cond-mat/9710223).
\bibitem{fm1} M. Franz and A.J. Millis, \prb {\bf 58}, 14572 (1998).
\bibitem{levin1} Q. Chen, K. Levin and I. Kosztin, \prb {\bf 63}, 184519
(2001).
\bibitem{ftv1} M.~Franz, Z.~Te\v{s}anovi\'c and O. Vafek, 
\prb {\bf 66}, 054535 (2002).
\bibitem{pbf} T.~Pereg-Barnea and M.~Franz, \prb {\bf 68}, 180506(R) (2003).
\bibitem{Hoffman} J.~Hoffman, K.~McElroy, D.-H.~Lee, K.M~Lang, H.~Eisaki, 
S.~Uchida, J.C.~Davis, Science {\bf 297} 1148(2002).
\bibitem{McElroy}  K.~McElroy, R.~W.~Simmonds, J.E.~Hoffman, D.-H.~Lee, 
J.~Orenstein, H.~Eisaki, S.~Uchida \&  J.C.~Davis, Nature {\bf 422} 592(2003).
\bibitem{Howald} C. Howald, H. Eisaki, N. Kaneko, M. Greven, and 
A. Kapitulnik, \prb {\bf 67}, 014533 (2003).
\bibitem{Wang}Q.-H.~Wang and D.-H.~Lee, \prb
{\bf 67}, 020511(2003).
\bibitem{Capriotti} L.~Capriotti, D.J.~Scalapino and R.D.~Sedgewick, 
\prb {\bf 68}, 014508 (2003).
\bibitem{Polkov} A. Polkovnikov, S. Sachdev and M. Vojta, Physica C 
{\bf 388}, 19 (2003).
\bibitem{Zhu}Lingyin Zhu, W. A. Atkinson, and P. J. Hirschfeld, 
\prb {\bf 69}, 060503 (2004).
\bibitem{Balatsky1} A.~V.~Balatsky, M.~I.~Salkola, and A.~Rosengren, 
\prb {\bf 51} 15547 (1995).
\bibitem{Pan1} S.~H.~Pan, J.~P.~O'Neal, R.~L.~Badzey,C.~Chamon,H.~Ding, 
J.~R.~Engelbrecht, Z.~Wang, 
H.~Eisaki,~S.~Uchida,~A.~K.~Gupta, K.~-W.~Ng, E.~W.~Hudson, K.~M.~Lang and 
J.~C.~Davis, 
Nature {\bf 413} 282 (2201).
\bibitem{Wang2}Ziqiang Wang, Jan R.~Engelbrecht, Shancai Wang, Hong Ding, and 
Shuheng Pan, \prb {\bf 65} 064509 (2002). 
\bibitem{note1} In the Born approximation the power spectrum is proportional 
to the  modulus square of the potential, see for example 
Ref. \onlinecite{Capriotti}.
\bibitem{peskin} See e.g.\ M. E. Peskin and D. V. Schroeder, {\em An 
Introduction to Quantum Field Theory}, Addison-Wesley 1995.
\bibitem{khvesh2} D.V. Khveshchenko, \prb {\bf 65}, 235111 (2002).
\bibitem{rantner2} W. Rantner and X.-G. Wen, \prl {\bf 86}, 3871 (2001).
\bibitem{gusynin1} V.P. Gusynin, D.V. Khveshchenko and  M. Reenders, 
\prb {\bf 67}, 115201 (2003). 
\bibitem{bena1} C. Bena, S. Chakravarty, J. Hu and C. Nayak, 
\prb {\bf 69}, 134517 (2004).
\bibitem{vershinin1} M. Vershinin, S. Misra, S. Ono, Y. Abe, Y. Ando, 
A. Yazdani, {\em Science} {\bf 303}, 1995 (2004).
\bibitem{hanaguri1} T. Hanaguri, C. Lupien, Y. Koshaka, D.-H. Lee, M. Azuma, 
M. Takano, H. Takagi, J.C. Davis, {\em Nature} (to appear).
\bibitem{fu1} H.C. Fu, J.C. Davis, D.H. Lee, cond-mat/0403001.
\bibitem{chen1} H.D. Chen, O. Vafek, A. Yazdani, S.-C. Zhang, 
\prl {\bf 93}, 187002 (2004).
\bibitem{tesanovic1} Z. Tesanovic, \prl {\bf 93}, 217004 (2004).
\bibitem{mcelroy1}  K. McElroy, D.-H. Lee, J.E. Hoffman, K.M. Lang, 
E.W. Hudson, H. Eisaki, S. Uchida, J. Lee, J.C. Davis, cond-mat/0404005.

\end{thebibliography}
\end{document}